\documentclass[12pt]{iopart}
\usepackage{iopams}
\usepackage[abbr,round]{harvard}
\usepackage{nicefrac}
\usepackage{graphicx,color,subfigure}
\usepackage{color}

\def\ga{\mathrel{\mathchoice {\vcenter{\offinterlineskip\halign{\hfil
$\displaystyle##$\hfil\cr>\cr\sim\cr}}}
{\vcenter{\offinterlineskip\halign{\hfil$\textstyle##$\hfil\cr
>\cr\sim\cr}}}
{\vcenter{\offinterlineskip\halign{\hfil$\scriptstyle##$\hfil\cr
>\cr\sim\cr}}}
{\vcenter{\offinterlineskip\halign{\hfil$\scriptscriptstyle##$\hfil\cr
>\cr\sim\cr}}}}}

\def\la{\mathrel{\mathchoice {\vcenter{\offinterlineskip\halign{\hfil
$\displaystyle##$\hfil\cr<\cr\sim\cr}}}
{\vcenter{\offinterlineskip\halign{\hfil$\textstyle##$\hfil\cr  
<\cr\sim\cr}}}
{\vcenter{\offinterlineskip\halign{\hfil$\scriptstyle##$\hfil\cr
<\cr\sim\cr}}}
{\vcenter{\offinterlineskip\halign{\hfil$\scriptscriptstyle##$\hfil\cr
<\cr\sim\cr}}}}}

\newcommand{\mur}{\mu_{\rm{r}}}
\newcommand{\Rm}{\rm{Rm}}
\newcommand{\vp}{{\varphi}}
\renewcommand{\vec}[1]{\mbox{\boldmath$#1$}}

\begin{document}

\title[High permeability disks in the Cadarache dynamo experiment]
{Influence of high permeability disks in an axisym-\\metric model of the
  Cadarache dynamo experiment}
\author{A. Giesecke$^1$, C. Nore$^{2,3}$, F. Stefani$^1$, G. Gerbeth$^1$,
  J. L\'eorat$^4$, W. Herreman$^{2,3}$, F. Luddens$^{2}$,  J.-L. Guermond$^{2,5}$}
\address{$^1$ Helmholtz-Zentrum Dresden-Rossendorf, 01328 Dresden, Germany}
\address{$^2$ Laboratoire d'Informatique pour la
      M\'ecanique et les Sciences de l'Ing\'enieur, CNRS, BP 133,
      91403 Orsay cedex, France}
\address{$^3$ Universit\'e Paris Sud 11, 91405 Orsay cedex, France}
\address{$^4$ Luth, Observatoire de Paris-Meudon, place Janssen, 92195-Meudon, France}
\address{$^5$ Department of Mathematics, Texas A\&M
      University 3368 TAMU, College Station, TX 77843, USA}
\ead{a.giesecke@hzdr.de}
\begin{abstract}
Numerical simulations of the kinematic induction equation are
performed on a model configuration of the Cadarache
von-K\'arm\'an-Sodium dynamo experiment.  The effect of a localized 
axisymmetric distribution of relative permeability $\mur$ that
represents soft iron material within the conducting fluid flow is
investigated.  
The critical magnetic Reynolds number
${\rm{Rm}}^{\rm{c}}$ for dynamo action of the first non-axisymmetric
mode roughly scales like 
${\rm{Rm}}_{\mur}^{\rm{c}}-{\rm{Rm}}_{\infty}^{\rm{c}} 
\propto \mu_r^{-1/2}$ 
i.e. the threshold decreases as $\mur$ increases. This scaling law
suggests a skin effect mechanism in the soft iron disks.  More
important with regard to the Cadarache dynamo experiment, we observe a
purely toroidal axisymmetric mode localized in the high permeability
disks which becomes dominant for large $\mur$. 
In this limit, the toroidal mode is close to the onset of dynamo
action with a (negative) growth-rate that is rather independent of the
magnetic Reynolds number. 
We qualitatively explain this effect by paramagnetic pumping at the
fluid/disk interface and propose a simplified model that
quantitatively reproduces numerical results. 
The crucial role of the high permeability disks for the mode
selection in the Cadarache dynamo experiment
cannot be inferred from computations using
idealized pseudo-vacuum boundary conditions 
($\vec{H}{\times}\vec{n} = 0$).
\end{abstract}
\submitto{\NJP}
\maketitle
\section{Introduction}
Astrophysical magnetic fields are a ubiquitous phenomenon.  They
affect formation and behavior of galaxies, stars or planets and might
even be important for structure formation on cosmic scales.  On
astrophysical scales fluid flow driven field generation by virtue of
the dynamo effect is relatively uncomplicated because the magnetic
Reynolds number is always huge.  However, due to their limited size
the realization of dynamo action in laboratory experiments is a
demanding task and requires an elaborate design that makes use of
optimizations like an ideal guidance of a fluid flow or a selective
choice of materials.
Material properties like electrical conductivity or relative
permeability have always been important for experimental dynamos. 
For example, the use of soft iron in the dynamo experiments of 
\citename{1963Natur.198.1158L}
\citeyear{1963Natur.198.1158L,1968Natur.219..717L}  
was crucial for the occur-\\rence of magnetic self excitation. 
More recently, the addition of high permeability mate-\\rial (soft
iron spheres) into a conducting fluid was examined to test magnetic
self excita-\\tion \cite{2003PhRvE..67e6309D} in a flow that otherwise
would not be able to sustain a dynamo.

The work presented here is motivated by the Cadarache
von-K\'arm\'an-Sodium (VKS) dynamo \cite{2007PhRvL..98d4502M}. 
In this experiment liquid sodium contained in a cylindrical vessel is
driven by two counter--rotating impellers that are located close to
the lids of the vessel.
Dynamo action is obtained only when (at least one of) the flow driving
impellers are made of soft iron with a relative permeability
$\mur\approx 65$ \cite{2010NJPh...12c3006V}. Moreover the
observed magnetic field is dominated by an axisymmetric mode
\cite{2009PhFl...21c5108M}.
It can be conjectured that the occurrence of dynamo
action with soft iron impellers and the axisymmetry of the magnetic field are
linked but, so far, the very nature of the axisymmetric dynamo is still
unknown.

A well-known necessary condition for the occurrence of dynamo action is
a sufficiently complex conducting fluid flow that couples the
toroidal and poloidal components of the magnetic field\footnote{In the
following, toroidal and poloidal components always refer to the axisymmetric
case so that the toroidal component corresponds to the azimuthal field
$\vec{B}_{\rm{tor}}=B_\vp\vec{e}_\vp$ and the poloidal component is given by
$\vec{B}_{\rm{pol}}=B_r\vec{e}_r+B_z\vec{e}_z$, where
$(\vec{e}_r,\vec{e}_\vp,\vec{e}_z)$ are the cylindrical unit vectors.}.
The interaction between these components gives rise to the so-called
dynamo cycle which consists of regenerating the toroidal field from
the poloidal field and vice versa.
This coupling can take place on large scales
\cite{1989RSPSA.425..407D} as well as on small scales by virtue of
the well known $\alpha$-effect \cite{1980mfmd.book.....K}.
It is less well known that non-homogeneities in the electrical conductivity
can also introduce such coupling and, by this, favour dynamo action. 
For example, a uniform flow over an infinite plate with varying
conductivity can produce dynamo action
\cite{1992gafd...64..135B,1993spd..conf..329W}. 
It is likely that inhomogeneous magnetic permeability can lead to
dynamo action as well.

In this paper we investigate the impact of a localized disk-like
permeability distribution embedded in a conducting axisymmetric
fluid flow on the growth-rates of the first axisymmetric and
non-axisymmetric magnetic eigenmodes. 
Induction effects due to non-axisymmetric perturbations (turbulence,
small-scale or large-scale flow or conductivity/permeability
distributions) are not taken into account.
First, we briefly re-examine the threshold of the equatorial dipole mode
as in~\citeasnoun{2010GApFD.104..505G} and propose an explanation for
the scaling law that relates the critical magnetic Reynolds
number to the permeability in the impeller disks.
Second we investigate the influence of the concentrated high
permeability on the axisymmetric field modes.
Even though they are always damped, according to Cowling's theorem
\cite{1933MNRAS..94...39C,1982GApFD..19..301H}, 
for large $\mur$ we find a dominant toroidal mode very close to the
onset of dynamo action.  
We suggest that this eigenmode plays a significant role in the
dominance of the axisymmetric mode in the dynamo observed in the
VKS experiment~\cite{2007PhRvL..98d4502M}.
\section{Model}\label{Sec:Model} 
\begin{figure}[t!]
\begin{center}
\includegraphics[width=9cm]{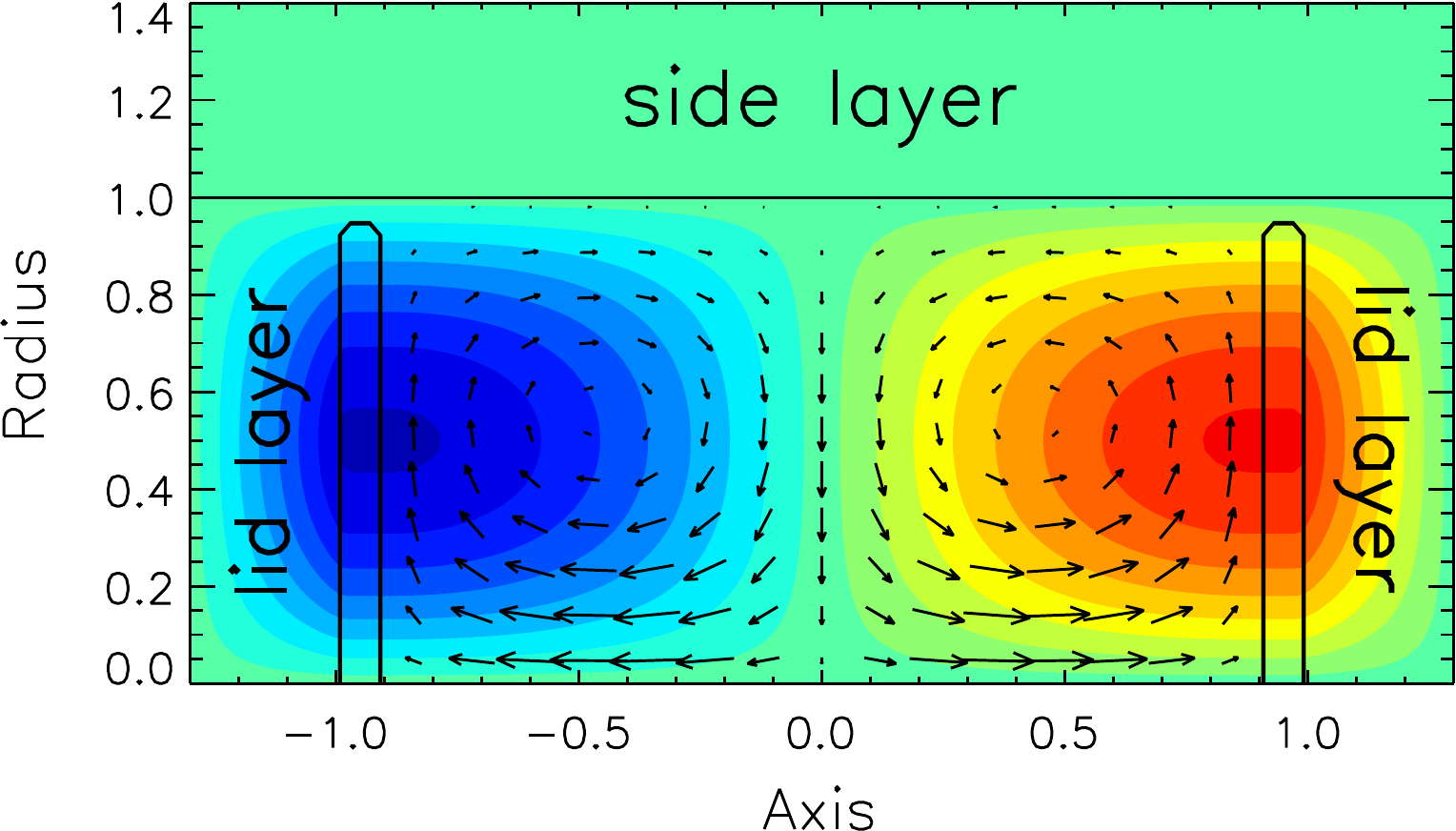}
\end{center}
\caption{Velocity field used in the kinematic simulations. The 
  azimuthal velocity, $u_\vp$, is shown in color and the poloidal
  component of the velocity, $u_r\vec{e}_r+u_z\vec{e}_z$, is shown
  with arrows.  
  The black structures in the intervals
  $z\in [-1.0;-0.9]$ and $z\in [0.9;1.0]$ represent two axisymmetric
  impellers of relative permeability $\mur > 1$. Note that the
  flow is mixed (poloidal and toroidal) 
  between the two impeller disks and purely toroidal within and
  behind the impeller disks.
  \label{fig::flowfield}}
\end{figure}

The induction equation with nonuniform material
coefficients, i.e. spatially dependent electrical conductivity
$\sigma=\sigma(\vec{r})$ and 
relative permeability $\mur=\mur(\vec{r})$, reads:
\begin{equation}
\frac{\partial\vec{B}}{\partial t}
= \nabla\!\times\!\left(\!\vec{u}\!\times\!\vec{B}
+\frac{1}{\mur\mu_0\sigma}\frac{\nabla\mur}{\mur}\!\times\!\vec{B}
-\frac{1}{\mur\mu_0\sigma}\!\nabla\!\times\!\vec{B}\!\right),\label{eq::indeq}
\end{equation}
where $\vec{u}$ is the prescribed (mean) flow, $\vec{B}$ the
magnetic flux density and $\mu_0$ the vacuum permeability
($\mu_0=4\pi\times 10^{-7}\mbox{Vs/Am}$).
The middle term in the right hand side of Eq.~(\ref{eq::indeq})
proportional to $\sim\nabla\mur\times\vec{B}$ represents the
so-called "paramagnetic pumping"~\cite{2003PhRvE..67e6309D}. 
This term is responsible for the suction of the magnetic field into
  the regions with large permeability and involves a (non divergence
  free) velocity-like field that we henceforth call "pumping velocity"
\begin{equation}
\vec{V}^{\mu}=\frac{1}{\mur\mu_0\sigma}\frac{\nabla\mur}{\mur}.
\label{eq::pump}
\end{equation}
The eigenvalue problem associated with equation~(\ref{eq::indeq}) is
solved numerically by using two different methods. One is based on a
spectral/finite element approach described in \citeasnoun{GLLNR11} 
(SFEMaNS, spectral/finite element method for Maxwell and Navier-Stokes
equations) which solves the eigenvalue problem using ARPACK.
The second approach utilizes a combined finite volume/boundary element method
for timestepping equation~(\ref{eq::indeq}), \citeasnoun{2008giesecke_maghyd}.
Both approximation methods can account for insulating boundaries and
non-uniform permeability and/or conductivity distributions.
In the FV/BEM scheme insulating boundary conditions are treated by
solving an integral equation 
on the boundary which allows a direct computation of the (unknown)
tangential field components by correlating the (known) normal field
components on the surface of the computational domain
\cite{iskakov2004,2008giesecke_maghyd}.  
In the SFEMaNS code the  magnetic field is
computed numerically in a certain domain outside of the cylinder and
matching conditions are used at the interfaces with the insulator
\cite{guermond2007}.
\newpage
The respective discretizations are done so that the transmission conditions
across the material interfaces are satisfied\footnote{At the
interface between two materials denoted 1 and 2, the transmission conditions
on the magnetic field and the electric
field/current are given by:
\\[0.2cm]
\begin{math}
\begin{array}{rclcrcll}
\vec{n}\cdot(\vec{B}^{\rm{1}}-\vec{B}^{\rm{2}}) & = & 0
&\mbox{ and } & \displaystyle
\vec{n}\times\left(\frac{\vec{B}^{\rm{1}}}{\mu_{{\rm{r}}}^{\rm{1}}}
-\frac{\vec{B}^{\rm{2}}}{\mu_{\rm{r}}^{\rm{2}}}\right) &=& 0 & 
\mbox{ for permeability jumps and} 
\\[0.3cm]
\vec{n}\cdot(\vec{j}^{\rm{1}}-\vec{j}^{\rm{2}})&=&0
&\mbox{ and }&
\vec{n}\times(\vec{E}^{\rm{1}}-\vec{E}^{\rm{2}})&=& 0 
&\mbox{ for conductivity jumps}.
\label{eq::jumpconditions_b1}
\end{array}
\end{math}
\\[0.1cm]
\noindent
where $\vec{n}$ denotes the unit normal vector at the
interface between both materials, $\vec{j}$ the current density and
$\vec{E}$ the electrical field.}.
In addition to having passed independent convergence tests on
manufactured solutions, the two codes have been cross-validated by
comparing their outputs on various common test cases (see 
\citename{andregafd}, \citeyear{andregafd},
\citename{2010GApFD.104..505G}, \citeyear{2010GApFD.104..505G}
and~\tref{tab::gr} below).
\\
We use the same configuration as applied in
\citeasnoun{PhysRevLett.104.044503}.
The computational domain is inspired from the VKS configuration.  The
conducting fluid is contained in a cylinder of height $H=2.6$ and
radius $R_{\rm{out}}=1.4$, surrounded by an insulator.
The fluid fills two unconnected compartments. 
The moving fluid is contained in an inner
cylinder of radius $R_{\rm{in}}=1$.
The fluid contained in the annular region comprised between the cylinders of
radius $R_{\rm{in}}=1$ and $R_{\rm{out}}=1.4$ is at rest; this region
is referred to as the side layer (see figure~\ref{fig::flowfield}).  
Two discoidal subdomains of radius
$R_{\rm{imp}}=0.95$ are located in the intervals $z\in [-1.0;-0.9]$
and $z\in [0.9;1.0]$ and are meant to model soft iron impeller disks of
thickness $d=0.1$; the relative magnetic permeability in these
two disks is denoted $\mur$.  
The velocity field $\vec{u}$ and the permeability distribution $\mur$
are assumed to be axisymmetric.
The velocity field between the impeller disks in the inner cylinder is
given by the so-called MND fluid flow \cite{2004phfl}:
\newpage
\begin{eqnarray}
u_r(r,z)&=&-(\pi/h) \cos\!\left({{2\pi z/h}}\right)r(1-r)^2(1+2r),\nonumber\\
u_{\varphi}(r,z)&=&4\epsilon r(1-r)\sin\left({{\pi z/h}}\right),\label{eq::s2t2}\\
u_z(r,z)&=&(1-r)(1+r-5r^2)\sin\left({{2\pi z/h}}\right),\nonumber
\end{eqnarray}
where $h$ is the distance between the two
impeller disks ($h=1.8$) and $\epsilon$ parametrizes the
toroidal to poloidal ratio of the flow (in the following we choose
$\epsilon=0.7259$)
\footnote{
This value is close to the optimum relation
between poloidal and toroidal flow that has been estimated in various 
comparable configurations \cite{ravelet2005} and has frequently
been utilized in previous studies of dynamo action driven by the
MND flow (e.g. \citename{2005physics..11149S}
\citeyear{2005physics..11149S},
\citename{2008giss} \citeyear{2008giss},
\citename{2010GApFD.104..505G} \citeyear{2010GApFD.104..505G}).
}.
A purely azimuthal velocity equal to the azimuthal velocity of the
MND flow at $z=\pm h/2$ is assumed in the two impeller disks.
A so-called lid layer \cite{2005physics..11149S} is added behind each
impeller disk, and the 
velocity field therein is modeled by linear interpolation along
the $z$-axis between the azimuthal velocity at the outer side of the
impeller disk and the no-slip condition at the lid of the vessel.
The velocity field and the impeller disks (two thin structures shown
in black solid lines) are displayed in~\fref{fig::flowfield}.
The conductivity is assumed to be uniform in the liquid metal and the
soft iron disks. 
We focus in this paper on non-uniform permeability distributions only,
which seems roughly justified for soft iron disks embedded in liquid
sodium.

The equations are nondimensionalized so that
$\mathcal{R}=R_{\rm{in}}$ is the reference length-scale
($R_{\rm{in}}$ is the radius of the flow active region) and
$\mathcal{U}=\max[(u_r^2+u_\vp^2+u_z^2)^{\nicefrac{1}{2}}]$ is the
reference velocity scale (maximum absolute value of the velocity
field). The control parameter is the magnetic Reynolds number
defined by ${\rm{Rm}}=\mu_0\sigma \mathcal{U}\mathcal{R}$. 
\section{Results}
The eigenvalues of the differential operator in the right-hand side
of equation (\ref{eq::indeq}) are denoted $\lambda=\gamma+i\omega$;
the real part $\gamma$ is the growth-rate of the field amplitude
($\gamma<0$ corresponds to decay) and the imaginary part $\omega$ is
the frequency.  All the computations reported below give
non-oscillatory eigen-modes (i.e. $\omega=0$).  
An immediate consequence of the axisymmetric 
setup is that the eigenmodes of the kinematic dynamo problem can be
computed for each azimuthal wavenumber $m$. 
\subsection{Overview}
\begin{table}[b!]
  \caption{Comparison of growth-rates obtained with FV (hybrid finite
    volume/boundary 
    element method) and SFEMaNS (spectral/finite element method for
    Maxwell and Navier-Stokes equations).  
    $\rm{Rm}$ is the magnetic Reynolds number, $\mur$ the disk permeability, 
    $\gamma_{m0}^{\rm{t}}$ the growth-rate of the axisymmetric toroidal field, 
    $\gamma_{m0}$
    the growth-rate of the axisymmetric mixed field,  
    $\gamma_{m1}$ the growth-rate of the first non-axisymmetric field
    ($m$1-mode). The thickness of the impeller disks is $d=0.1$.}\label{tab::gr}
\begin{center}
\begin{tabular}{rrrrrl}
\hline
${\rm{Rm}}$  & $\mur$ & $\gamma_{m0}^{\rm{t}}$ & $\gamma_{m0}$ &
$\gamma_{m1}$ & {\rm{Scheme}}\\
\hline
  0 &  1 & -8.950 & -4.159 & -4.273 & FV\\
  0 &  1 & -8.977 & -4.162 & -4.322  & SFEMaNS\\
  0 & 60 & -1.292 & -3.887 & -1.715 & FV\\
  0 & 60 & -1.305 & -3.893 & -1.722 & SFEMaNS\\
\hline
 30 &  1 & -8.748 & -3.591 & -2.690 & FV\\
 30 &  1 & -8.770 & -3.597 & -2.780 & SFEMaNS\\
 30 & 60 & -1.134 & -3.404 & -2.511 & FV\\
 30 & 60 & -1.155 & -3.478 & -2.476 & SFEMaNS\\
\hline
 70 &  1 & -8.079 & -3.467 & -0.119 & FV\\
 70 &  1 & -8.119 & -3.471 & -0.215 & SFEMaNS\\
 70 & 60 & -1.203 & -3.232 &  1.012 & FV\\
 70 & 60 & -1.219 & -3.264 &  0.969 & SFEMaNS\\
\hline
\hline
\end{tabular}
\end{center}
\end{table}
We show in \tref{tab::gr} sample values of growth-rates obtained by
FV and SFEMaNS for the above simplified VKS model problem. 
This table confirms that FV and SFEMaNS converge to the same solutions up
to $2\%$ on the growth-rates.  
We use the following notation
in~\tref{tab::gr} and in the rest of the paper:
$\gamma_{m0}$ is the growth-rate of a mixed poloidal/toroidal mode.  
This mode degenerates to a purely poloidal mode when ${\rm{Rm}}=0$, and
when there is no permeability jump (e.g. for stainless steel disks) this
mode always determines the behavior of the axisymmetric field.  
Furthermore, $\gamma_{m0}^{\rm{t}}$ is the growth-rate of the first
axisymmetric mode ($m=0$) that is purely toroidal and
$\gamma_{m1}$ is the growth-rate of the first non-axisymmetric mode
($m=1$). 

\begin{figure}[b!]
\centerline{
\subfigure[${\rm{Rm}}=30$, $\mur=1$, $m=1$, decay]{
\includegraphics[width=5cm,angle=90]{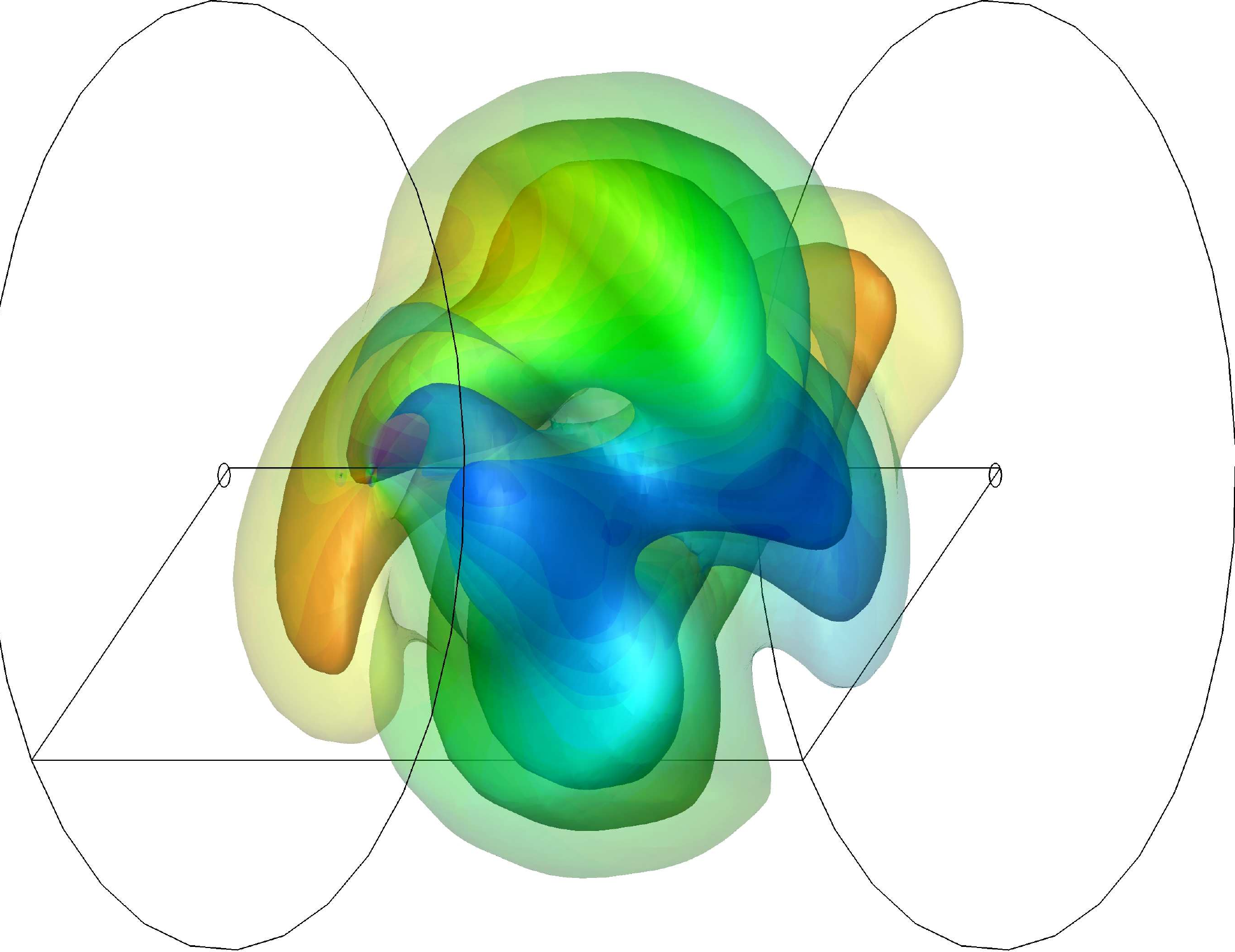}}
\subfigure[${\rm{Rm}}=30$, $\mur=60$, $m=0$, decay]{
\includegraphics[width=5cm,angle=90]{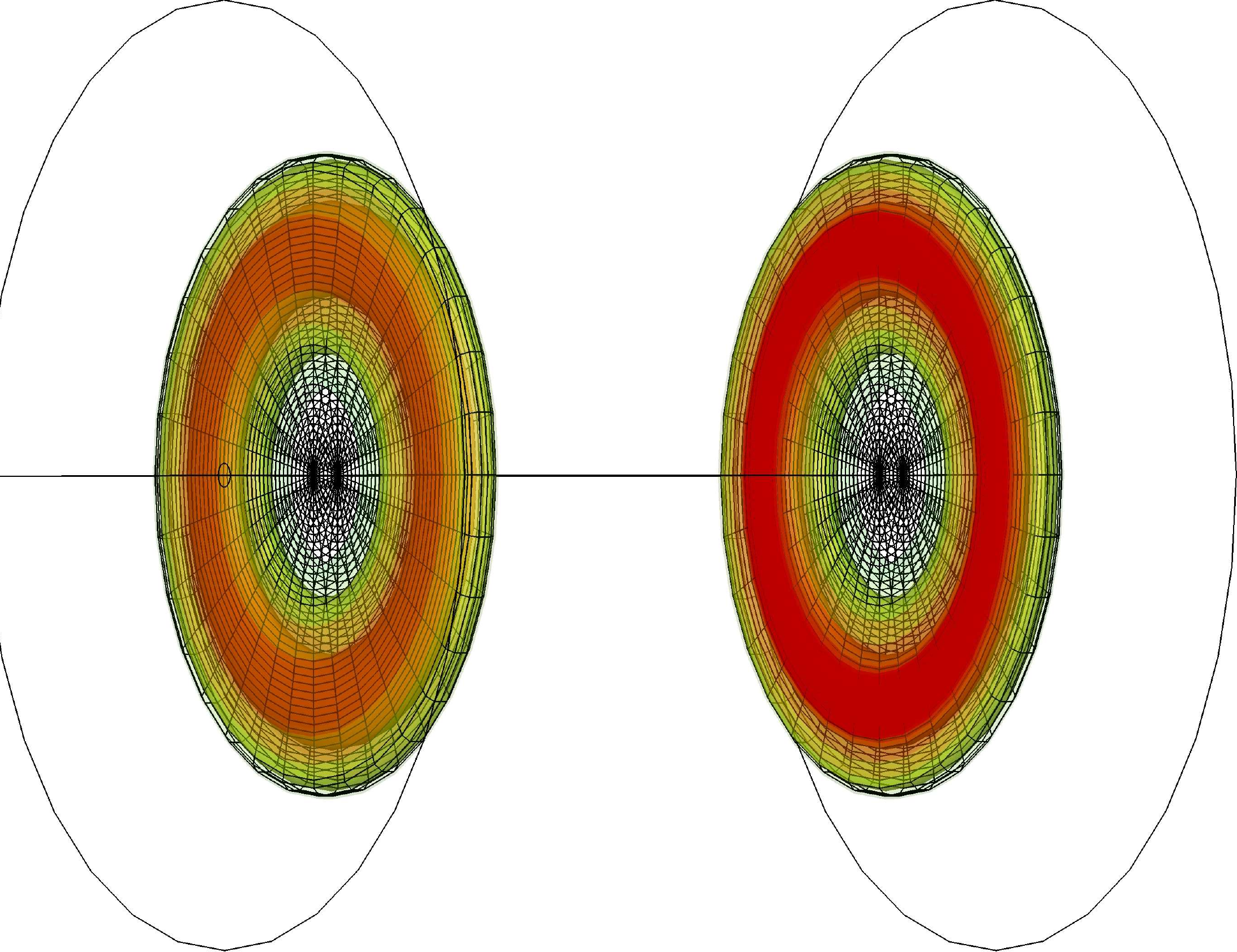}}
}
\vspace*{0.2cm}
\centerline{
\subfigure[${\rm{Rm}}=70$, $\mur=1$, $m=1$, decay]{
\includegraphics[width=5cm,angle=90]{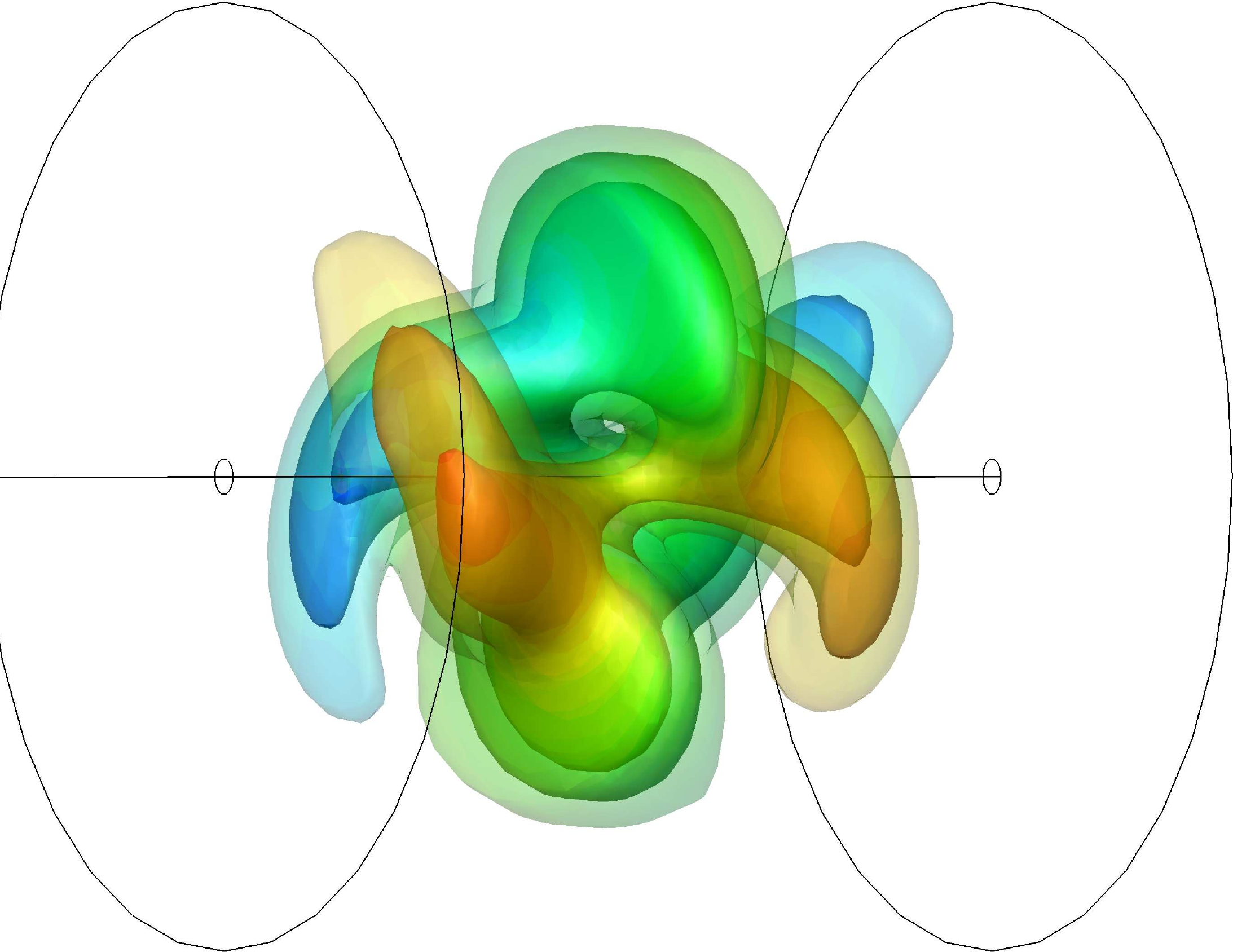}}
\subfigure[${\rm{Rm}}=70$, $\mur=60$, $m=1$, growth]{
\includegraphics[width=5cm,angle=90]{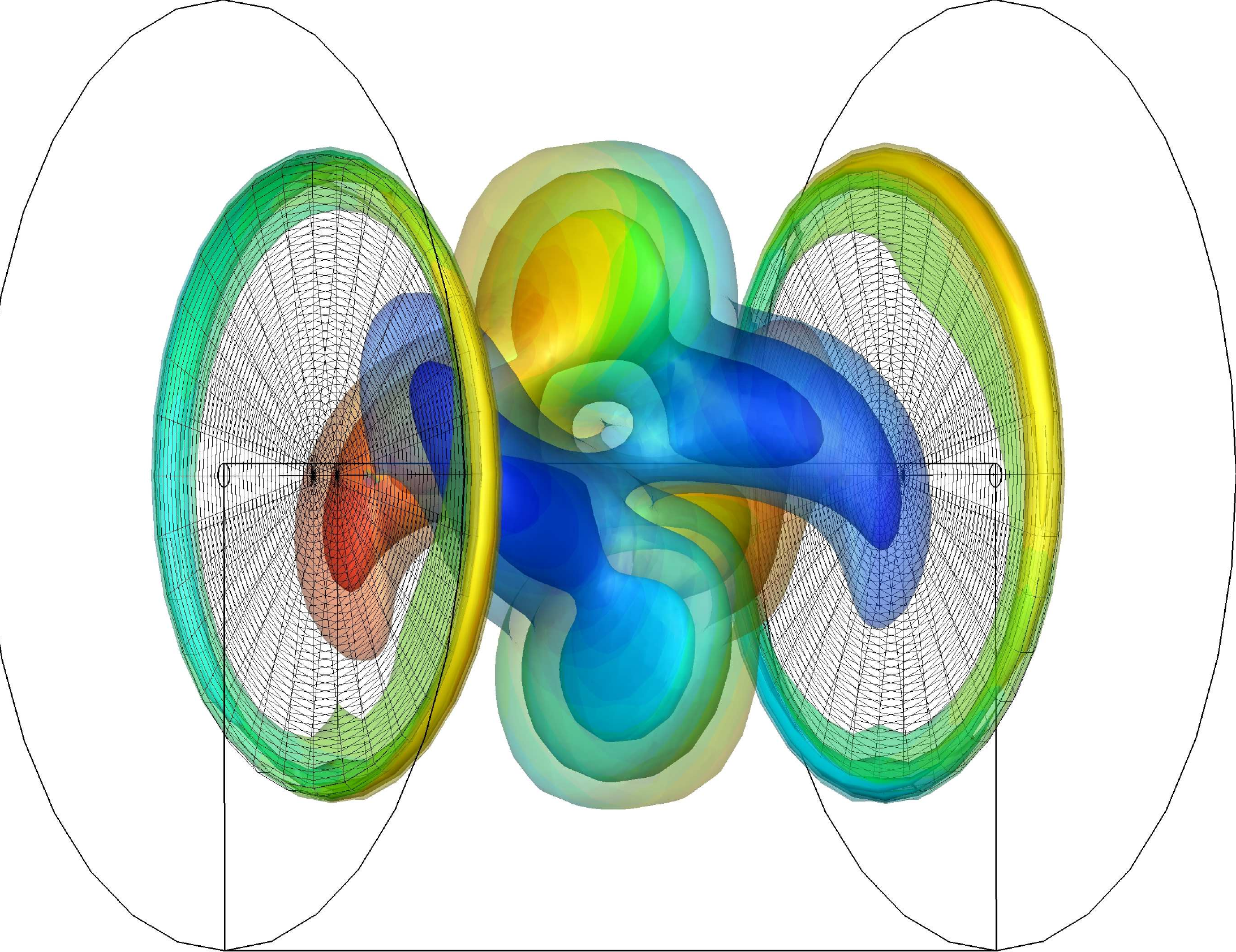}
}
}
\caption{Dominant magnetic eigenmodes. The isosurfaces
  show the magnetic energy density at $10\%, 
  20\%, 40\%$ of the maximum value. The colors show 
  $H_\vp=\mur^{-1}B_\vp$
  and the meshes show the location of the soft iron disks (in the
  right column). Note that all eigenmodes are decaying except
  the case ${\rm{Rm}}=70, \mur=60$ (lower
  right panel) which is above the dynamo threshold and therefore
  presents a growing $m1$--mode.\label{fig::structure_noblades}} 
\end{figure}

When $\Rm = 0$, the dominant and the second dominant $m0$-modes at
$\mur=1$ are purely poloidal and purely toroidal, respectively; the
situation is reversed at $\mur=60$: the dominant mode is purely
toroidal. All growth-rates increase with $\mur$.
When $\Rm > 0$, the growth-rate of the $m0$-mode is always negative in
agreement with Cowling's 
theorem~\cite{1933MNRAS..94...39C}, 
but we observe that its relaxation time becomes longer as the
permeability in the impeller disks increases.  We also observe that
dynamo action occurs on the $m1$-mode and that increasing $\mu_r$
lowers the critical threshold on $\Rm$.

Snapshots of the dominant magnetic eigenmode 
are shown in~\fref{fig::structure_noblades}.
The structure of the $m1$-mode does not change very much
with respect to $\mur$ and ${\rm{Rm}}$;
it is an equatorial dipole with two opposite axial structures mainly
localized in the bulk of the fluid. 
In contrast, the $m0$-mode is essentially localized in the two
impeller disks and does barely differ from the pattern obtained in 
the free decay case (see~\fref{fig::structure_freedecay} below).
\subsection{Effect of the disk permeability on the $m1$-mode}
\begin{figure}[h!]
\begin{center}
\includegraphics[width=0.5\textwidth]{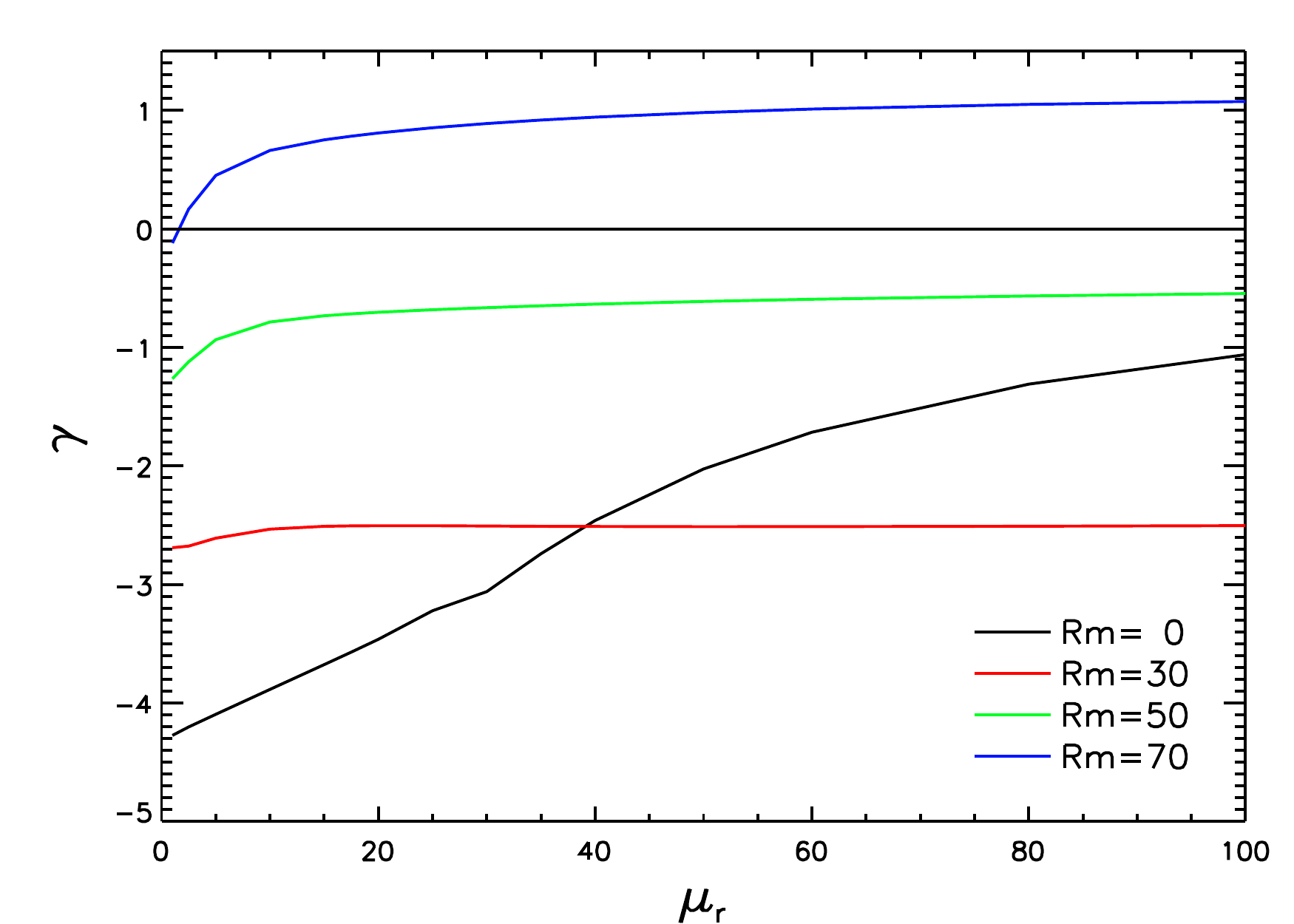}
\end{center}
\caption{Growth-rate of the $m1$-mode as a function of $\mur$ for
  ${\rm{Rm}}=0, 30, 50, 70$.}   
\label{fig::gr_vs_mur_m1} 
\end{figure}

\begin{figure}[h!]
\begin{center}
\includegraphics[width=0.5\textwidth]{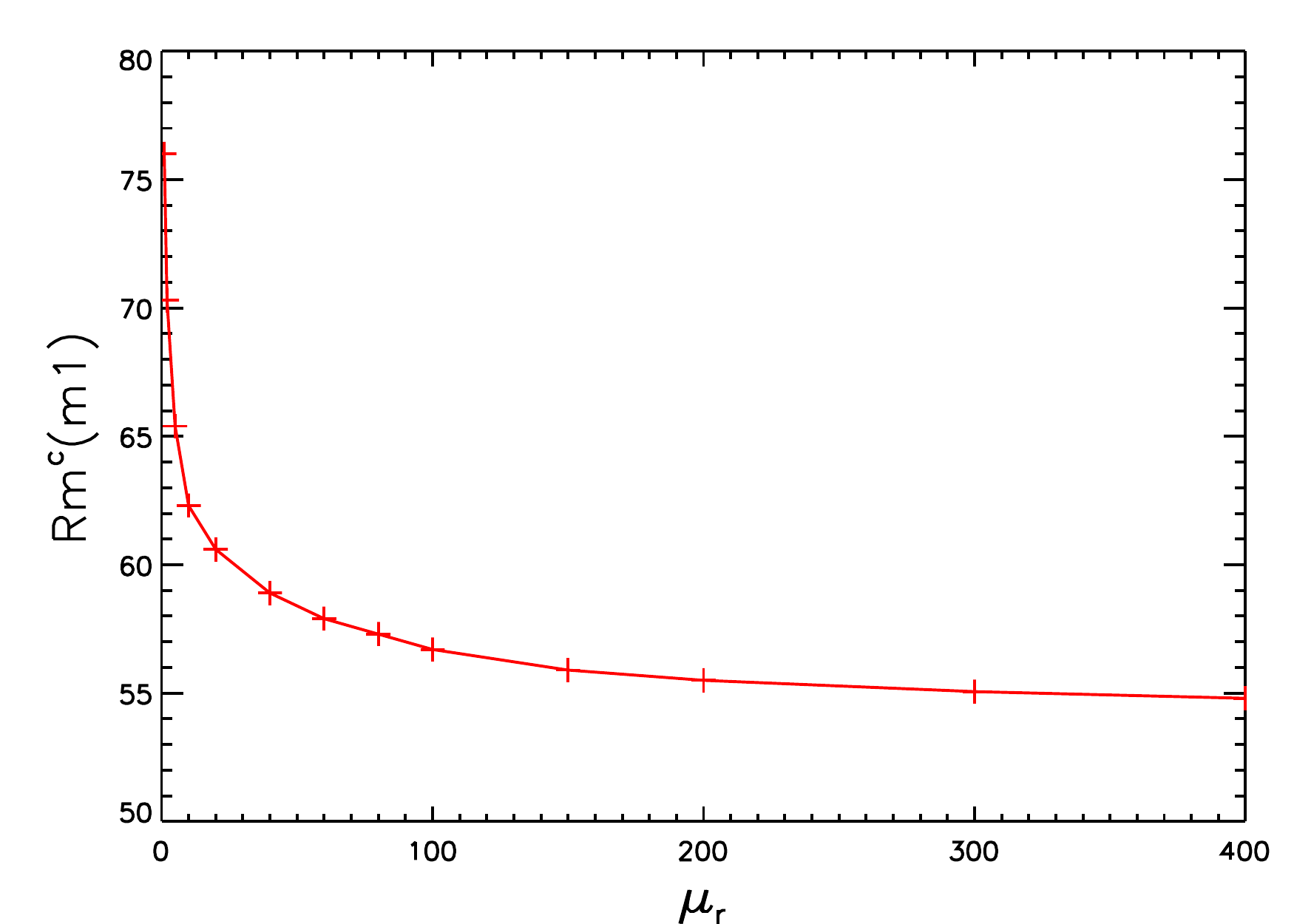}
\nolinebreak[4!]
\includegraphics[width=0.5\textwidth]{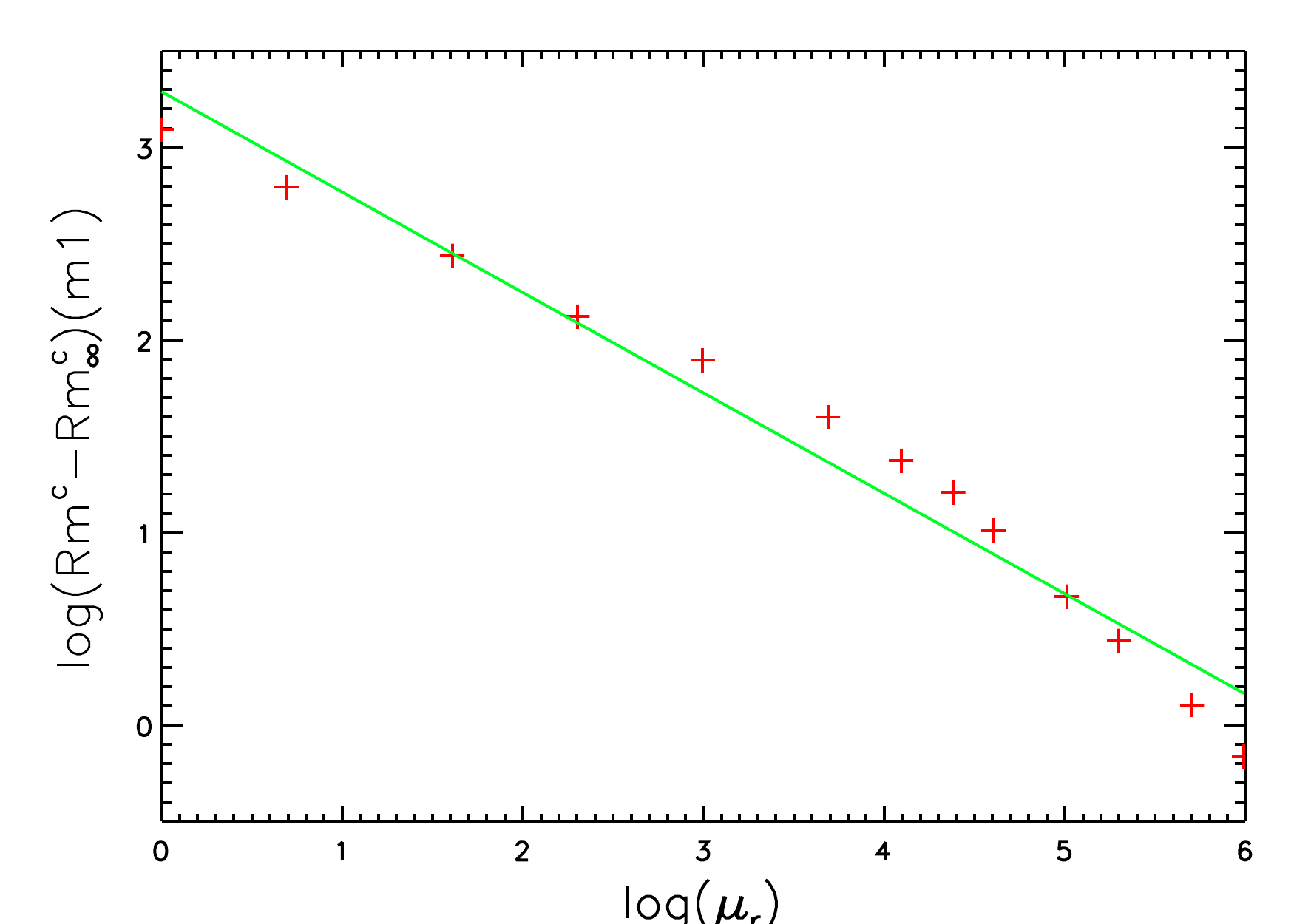}
\end{center}
\caption{ 
Left panel: Dynamo threshold for the $m1$-mode as a function of $\mu_r$.
  Right panel: Linear fit on $\log$-$\log$ scale provides the scaling 
  ${\rm{Rm}}^{\rm{c}}-{\rm{Rm}}^{c}_{\infty} \propto \mu_r^{-0.52}$.
\label{fig::rmc_vs_mur_m1}}
\end{figure}

Figure~\ref{fig::gr_vs_mur_m1} shows the growth-rate of the $m1$-mode
as a function of the relative permeability of the impeller disks for
four values of the Reynolds number.
This figure is similar to figure 13a in \citeasnoun{2010GApFD.104..505G}.
The growth-rate reaches quickly an asymptotic value 
when $\rm{Rm}>0$, which is not the case when $\rm{Rm}=0$.
The $m1$-mode clearly depends on ${\rm{Rm}}$ and exhibits
dynamo action     
when ${\rm{Rm}}$ is large enough.
It can be seen in the left panel in~\fref{fig::rmc_vs_mur_m1} that the
threshold for dynamo action goes from ${\rm{Rm}}^{\rm{c}} \approx
76$ when $\mur=1$ to ${\rm{Rm}}^{\rm{c}}_\infty \approx 53.95$ when
$\mur \rightarrow \infty$.
The asymptotic threshold for $\mur \rightarrow \infty$ has been
calculated by enforcing the boundary condition $\vec{H}{\times}
\vec{n} = 0$ on the impeller disks
(pseudo-vacuum or Vanishing Tangential Field
condition) while keeping the flow pattern~(\ref{eq::s2t2}) unchanged. 
This computation shows that, as far as the $m1$-mode is concerned, the
impeller disks behave like an idealized ferromagnetic material in the
limit $\mur \rightarrow \infty$. 
Upon inspection of the right panel in~\fref{fig::rmc_vs_mur_m1},
where ${\rm{Rm}}^{\rm{c}} - {\rm{Rm}}^{\rm{c}}_\infty$ is displayed
as a function of $\mu_r$ in $\log$-$\log$ scale, we infer the
following scaling law: ${\rm{Rm}}^{\rm{c}} -
{\rm{Rm}}^{\rm{c}}_\infty \propto \mu_r^{-0.52}$. This type of
scaling is an indication that a boundary layer effect is at play
which can be explained as follows.
Starting with the idea that the stationary $m1$-dynamo is generated
within the fluid flow, it is reasonable to expect that the rotating
disks see this field as a quickly varying imposed field.  The
magnetic field cannot penetrate the disks when the permeability is
infinite but, according to the classical skin effect, it can diffuse
through a thin boundary layer of thickness $\delta = (\sigma \mu_r
\Omega)^{-1/2}$ when $\mu_r$ is finite ($\Omega$ is the mean angular
velocity). This diffusion effect adds a supplementary $\mu_r^{-1/2}$
damping to the magnetic field mode compared to the infinite
permeability case. 
\subsection{Effect of the disk permeability on the $m0$-mode}
\begin{figure}[h!]
\begin{center}
\includegraphics[width=15cm]{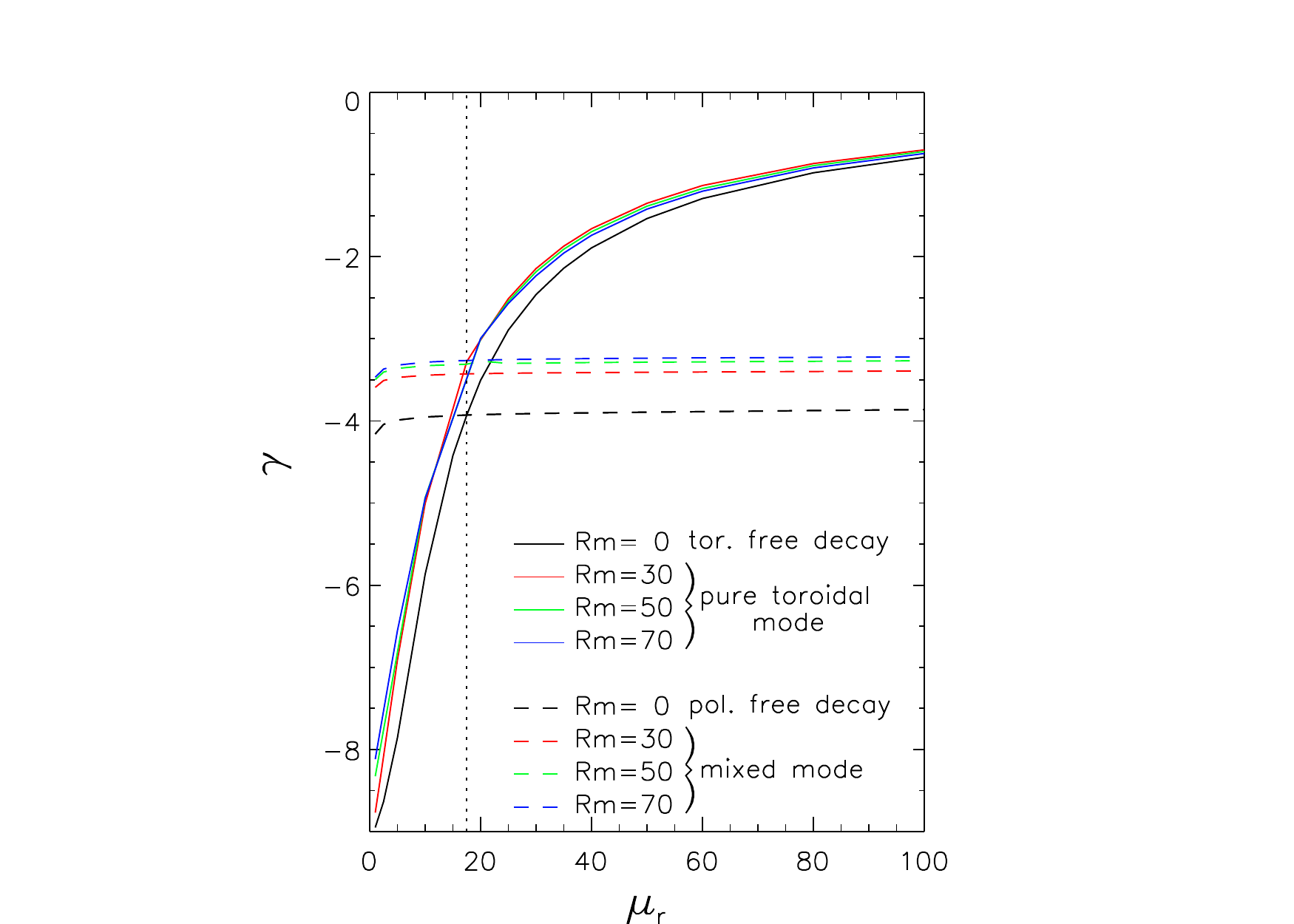}
\end{center}
\caption{
Growth-rates of the $m0$-mode. The dotted vertical line marks the
transitional value $\mur^{\rm{t}} \approx 17.5$ at which the pure
toroidal mode becomes dominant. \label{fig::gr_vs_mur_m0}} 
\end{figure}

Figure~\ref{fig::gr_vs_mur_m0} shows the growth-rates of the
axisymmetric mode as a function of $\mur$.
Contrary to what we have observed for the $m1$-mode, the dependence
of the $m0$-mode with respect to ${\rm{Rm}}$ seems to be
small; more precisely, the flow does not seem to play a significant
role when the permeability is large.

\begin{figure}[h!]
\begin{center}
\includegraphics[width=5cm,angle=90]{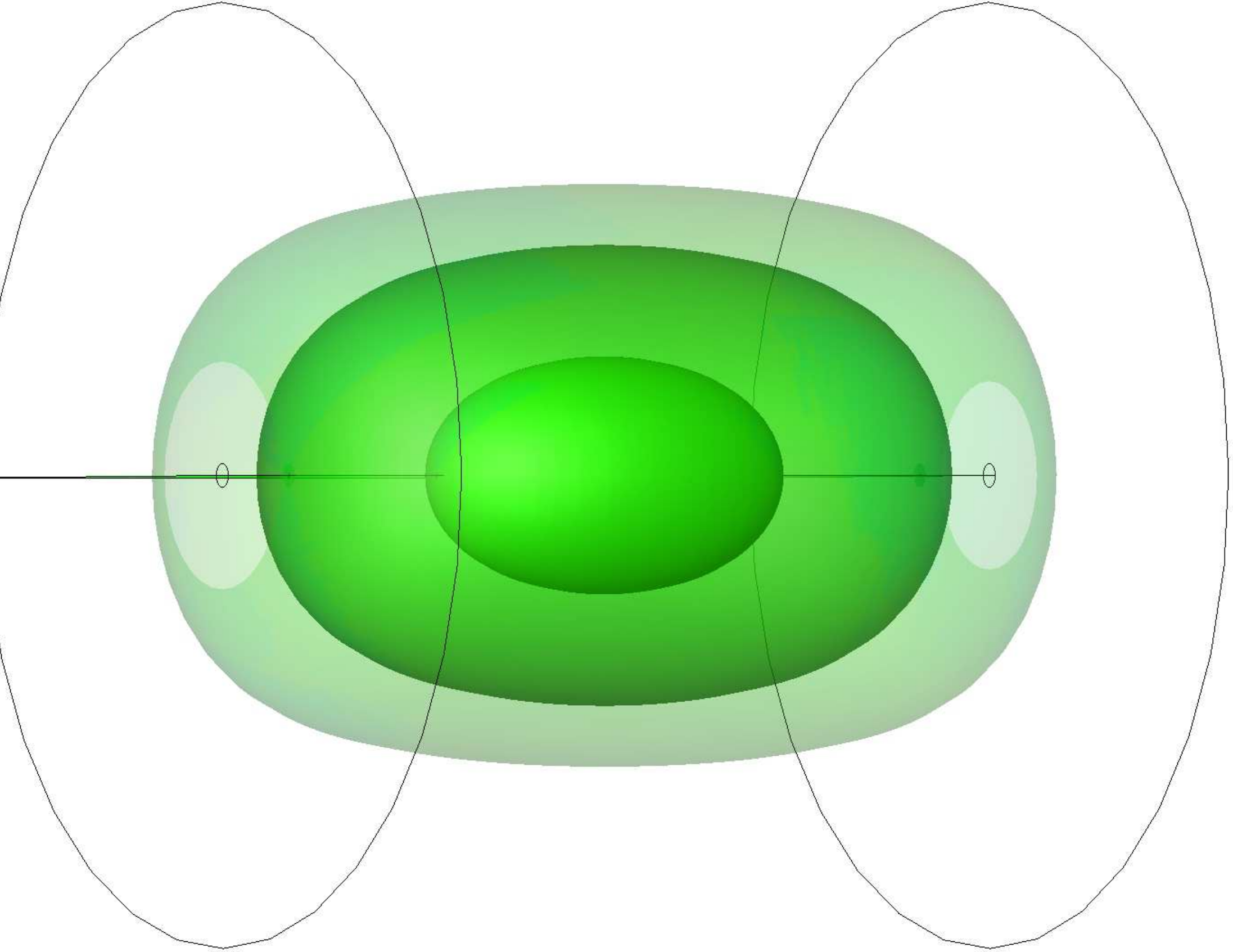}
\nolinebreak[4!]
\includegraphics[width=5cm,angle=90]{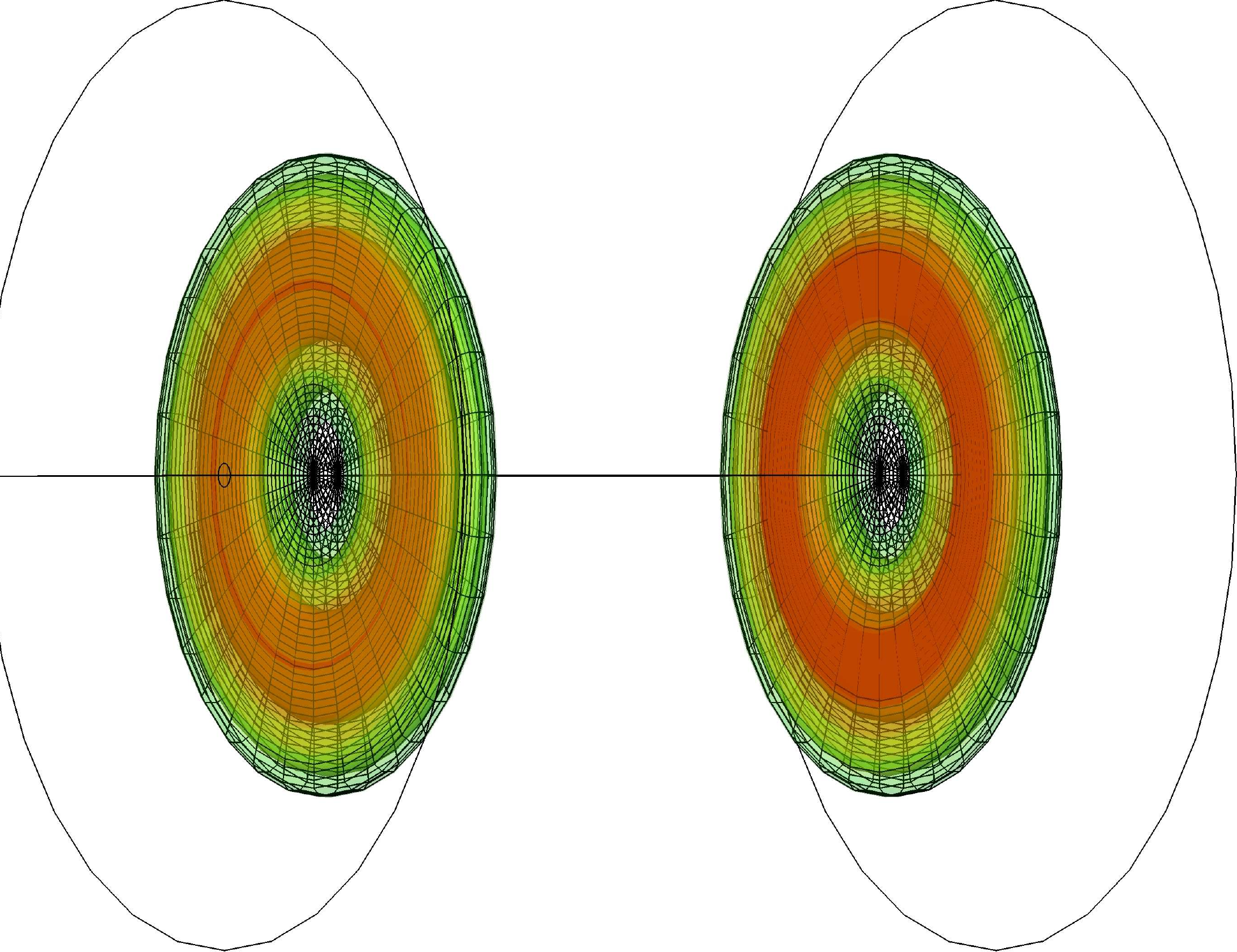}
\end{center}
\caption{
Spatial structure of $m0$-mode in free decay (${\rm{Rm}}=0$). Left
panel: $\mur=1$, right panel: 
$\mur=60$. The isosurfaces show the energy density at 20\%, 40\%,
80\% of the maximum value. The colors code the
azimuthal component $H_\vp=\mur^{-1}B_\vp$.
\label{fig::structure_freedecay}}
\end{figure}

In the free decay case (${\rm{Rm}}=0$) the poloidal (dashed black
line) and toroidal (solid black line) modes 
are decoupled and the growth-rates of these two modes are
$\gamma_{m0}=-4.159$ and $\gamma^{\rm{t}}_{m0}=-8.950$,
respectively.
The decay time of the poloidal eigenmode is significantly larger
than that of the toroidal one.
The dominant poloidal eigenmode exhibits a typical dipolar pattern as
shown in the left panel of~\fref{fig::structure_freedecay}.  
Increasing the disk permeability (still at ${\rm{Rm}}=0$) the poloidal mode
remains nearly unaffected (dashed black curve in~\fref{fig::gr_vs_mur_m0})
whereas the purely toroidal mode is significantly enhanced and eventually
becomes dominant when  
$\mur\ga \mur^{\rm{t}} \approx 17.5$ (see solid curves in
figure~\ref{fig::gr_vs_mur_m0} and right panel 
in~\fref{fig::structure_freedecay}).
The growth-rate of the toroidal mode increases monotonically with
$\mur$ and roughly scales $\propto -\mur^{-1}$.

The introduction of a velocity field (${\rm{Rm}}>0$) 
transfers poloidal field components into toroidal field components,
but not vice versa. Therefore for increasing ${\rm{Rm}}$ a
mixed mode is generated from the purely poloidal field that is 
observed at $\rm{Rm}=0$ 
(see dashed lines in~\fref{fig::gr_vs_mur_m0}).  
The dependence of the growth-rate of the mixed mode with respect to
the Reynolds number and the permeability is small.  
This mixed mode is dominant when $\mur\la\mur^{\rm{t}}$, but above this
transitional point it is the purely toroidal eigenmode that dominates 
(see solid colored curves in~\fref{fig::gr_vs_mur_m0}).
The purely toroidal mode hardly depends on the magnetic Reynolds number
and its growth-rate increases with $\mur$ like in the free decay situation.

Surprisingly, the value of the transitional permeability
$\mur^{\rm{t}}$ is more or less the same for all the
considered Reynolds numbers (see the vertical dotted line
in~\fref{fig::gr_vs_mur_m0} that marks the transition).
\section{Discussion on the $m0$-mode}
\subsection{The coupling}
Using the cylindrical coordinate system $(r,\vp,z)$, and assuming
  axisymmetry, the induction equation can be written as follows:
\begin{eqnarray}
\fl
\displaystyle \left ( \frac{\partial}{\partial t}
+ u_r \frac{\partial }{\partial r} + u_z \frac{\partial }{\partial z} \right)  B_{r} 
& = & \left (\!B_r \frac{\partial }{\partial r} 
\!+\! B_z \frac{\partial }{\partial z}\!\right)\! u_{r}
+\eta_0\!\left[
\frac{\partial}{\partial z}\!
\left(\frac{\partial}{\partial z}\frac{B_{r_{{}_{ }}}}{\mur}
-\frac{\partial}{\partial r}\frac{B_{z_{{}_{ }}}}{\mur}\right)
\right],
\label{eq::br}
\\[0.2cm]
\fl
\displaystyle \left [ \frac{\partial}{\partial t}
+ u_r\!\left ( \frac{\partial }{\partial r} - \frac{1}{r} \right ) 
\!+u_z \frac{\partial }{\partial z} \right]\!  B_{\vp} 
& = & \left [ B_r \left ( \frac{\partial }{\partial r} 
- \frac{1}{r} \right ) 
+ B_z \frac{\partial }{\partial z} \right]  u_{\vp}
+ \eta_0 \Delta_* \frac{B_{\vp}}{\mu_r},
\label{eq::bphi}
\\[0.2cm]
\fl
\displaystyle \left ( \frac{\partial}{\partial t}
+ u_r \frac{\partial }{\partial r} + u_z \frac{\partial }{\partial z} \right)  B_{z} 
& = & \left(\! B_r \frac{\partial }{\partial r} 
\!+\! B_z \frac{\partial }{\partial z}\!\right)  u_{z}
-\eta_0\frac{1}{r}
\frac{\partial}{\partial r}\!\left[\!r\!\left(\frac{\partial}{\partial z}
\frac{B_{r_{{}_{ }}}}{\mur}-\frac{\partial}{\partial r}
\frac{B_{z_{{}_{ }}}}{\mur}\right)\!\right]\label{eq::bz}
\end{eqnarray}
where $\Delta_* = \frac{\partial^2}{\partial r^2}
+ \frac{1}{r} \frac{\partial}{\partial r} +\frac{\partial^2}{\partial
  z^2} -\frac{1}{r^2}$, and $\eta_0=\frac{1}{\sigma
  \mu_0}$.
This form of the induction equation clearly shows that $B_r=B_z=0$ and
$B_\varphi \neq 0$ can be an axisymmetric solution: this is the purely
toroidal mode.   
If $B_r \neq 0, \, B_z \neq 0$, then their shearing by the nonzero
azimuthal flow $u_\varphi$,  the so-called $\Omega$-effect, will
always generate $B_\varphi \neq 0$ which then 
results in a mixed mode. The growth-rate of the mixed mode will
however remain entirely fixed by its poloidal components $B_r$ and
$B_z$ (see~\fref{fig::gr_vs_mur_m0}).

Purely toroidal and purely poloidal fields cannot exist
if $\mur$ depends on $\vp$. The same remark holds if $\sigma$
depends on $\vp$; for instance, spatial variation of the electric
conductivity is used in~\cite{1992gafd...64..135B} to produce dynamo
action. When the permeability $\mur$ is axisymmetric there is no
mechanism to transfer magnetic energy from the toroidal component
$B_\vp$ to the poloidal pair $(B_r,B_z)$ (see Eqs.~\ref{eq::br}
and~\ref{eq::bz}).
\subsection{Selective enhancement of $B_\vp$}
We start by explaining qualitatively why, for large values of $\mur$,
the purely toroidal mode is the 
least damped one and why the mixed mode is so little influenced by
the disk. The argument is based on the paramagnetic pumping term
(\ref{eq::pump}) that is at the origin of an electromotive force
(EMF): 
\begin{equation} \vec{\mathcal{{E}}}^{{\mu}}=
  \vec{V}^{\mu}{\times}\vec{B}=\frac{1}{\mu_0\mur\sigma}\frac{\nabla\mur}{\mur}{\times}\vec{B}.
\end{equation} 
Since the permeability jump is restricted to the material
interface there is only a contribution to the EMF within that
localized area.  
\begin{figure}[t!]
  \begin{center}
    \hspace*{1cm}
    \includegraphics[width=9cm]{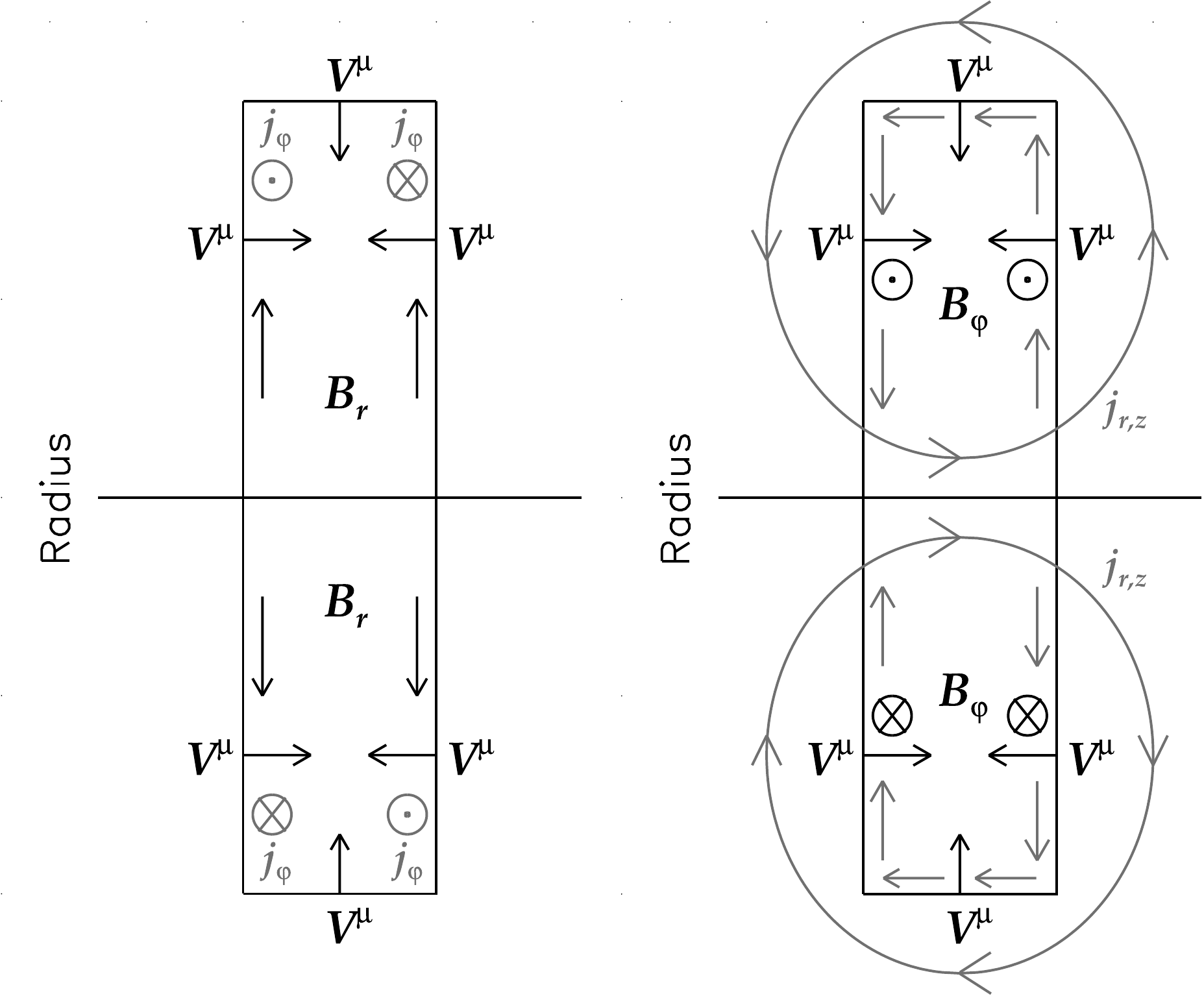}
  \end{center}
  \caption{Left panel: current generation at the fluid-disk interface by
    paramagnetic pumping for a radial magnetic field $B_r$.  Right panel: current
    generation from paramagnetic pumping for the azimuthal magnetic
    field $B_\vp$.\label{fig::emf_current_sketch}}
\end{figure}
For sufficiently thin disks, as considered here, it is reasonable to
assume that the permeability jump at the rim of the impeller disks
plays a minor role.  We therefore assume that the pumping velocity is
mainly axial: $\vec{V}^{\mu}\propto \frac{\partial}{\partial z}\mur
\vec{e}_z$. The interaction of $\vec{V}^{\mu}$ with the axial field
$B_z\vec{e}_z$ is henceforth neglected.

\begin{figure}[b!]
\begin{center}
\hspace*{-0.5cm}
\includegraphics[width=7cm]{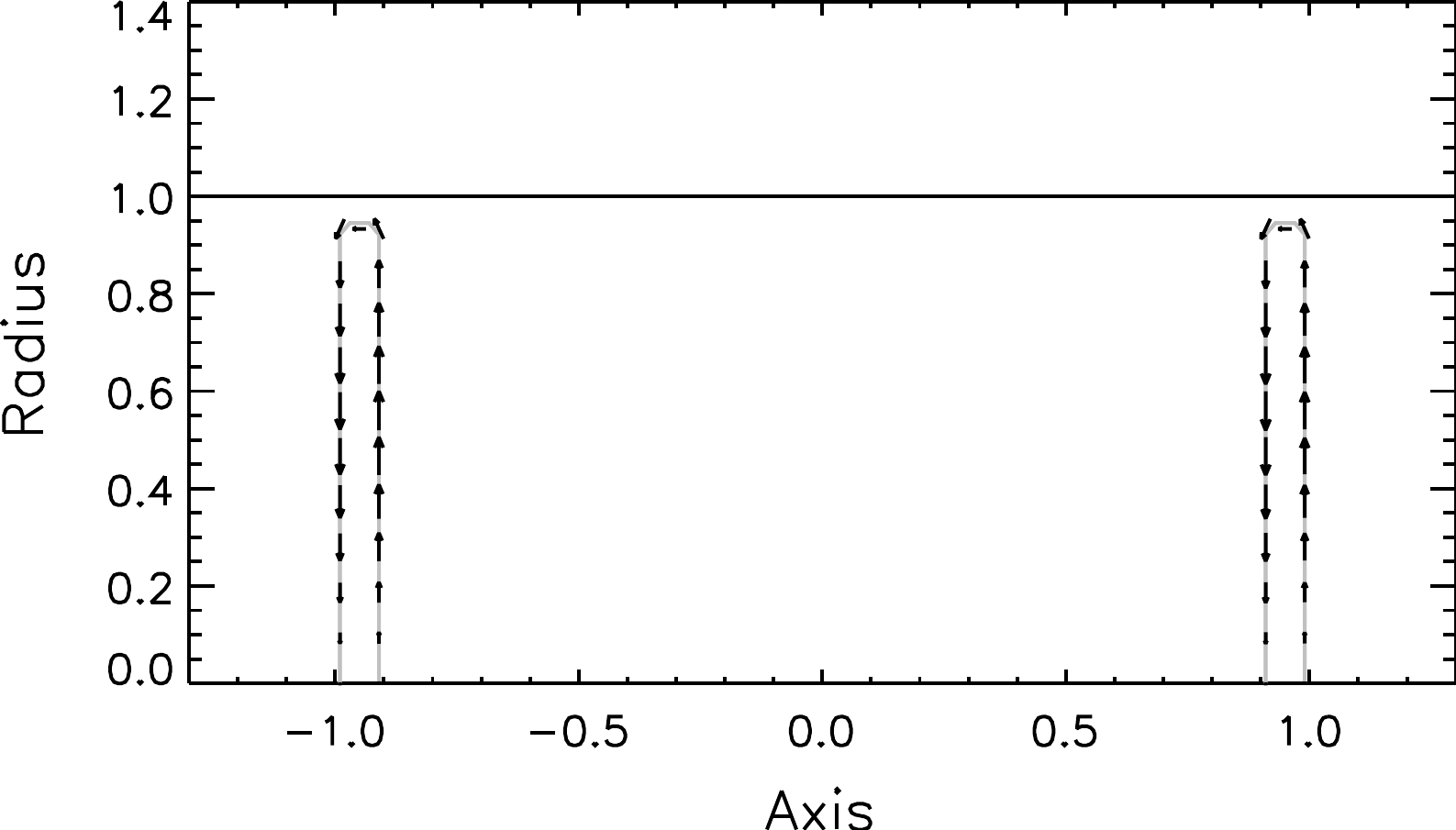}
\nolinebreak[4!]
\includegraphics[width=7cm]{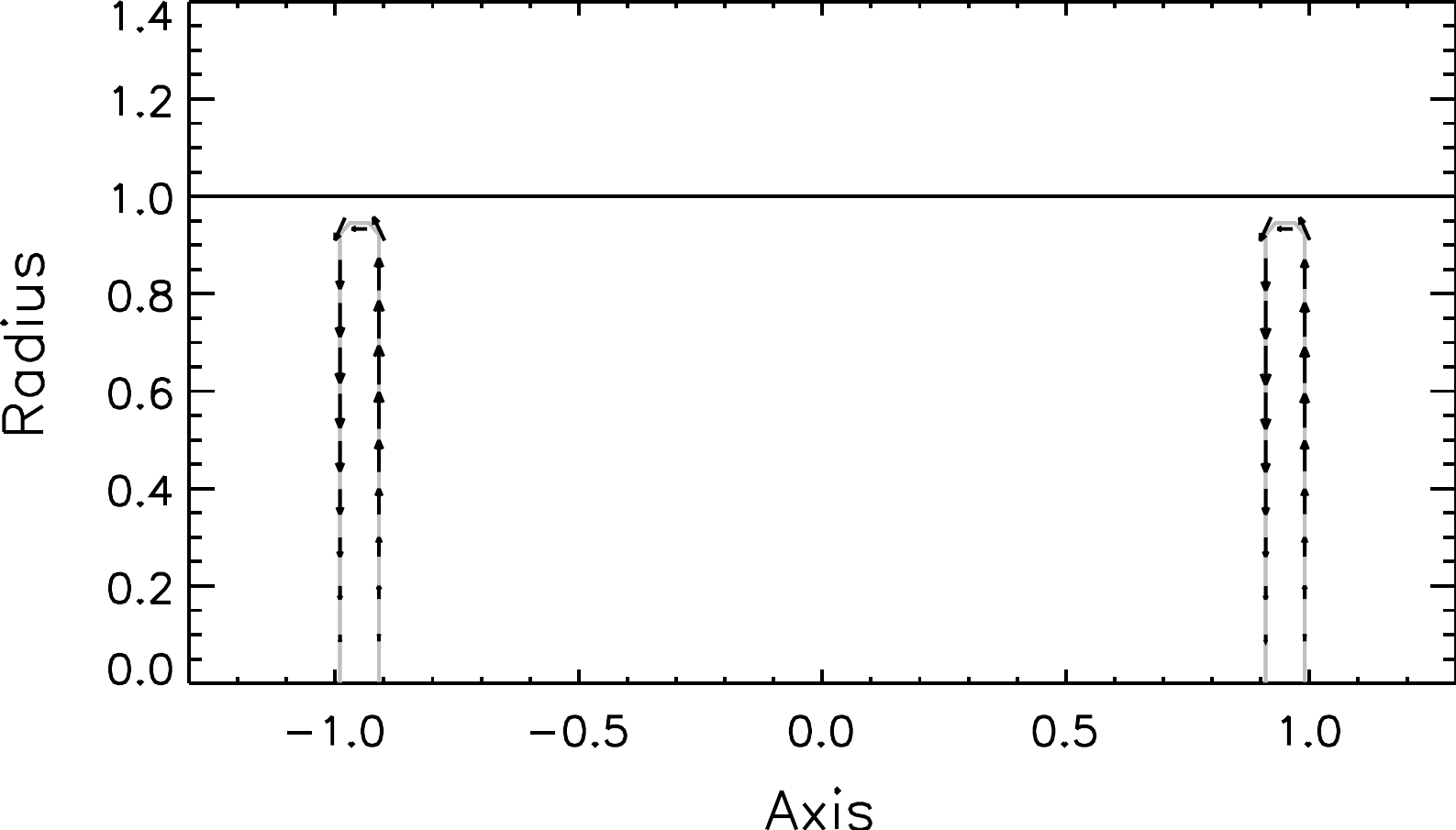}
\\
\hspace*{-0.5cm}
\includegraphics[width=7cm]{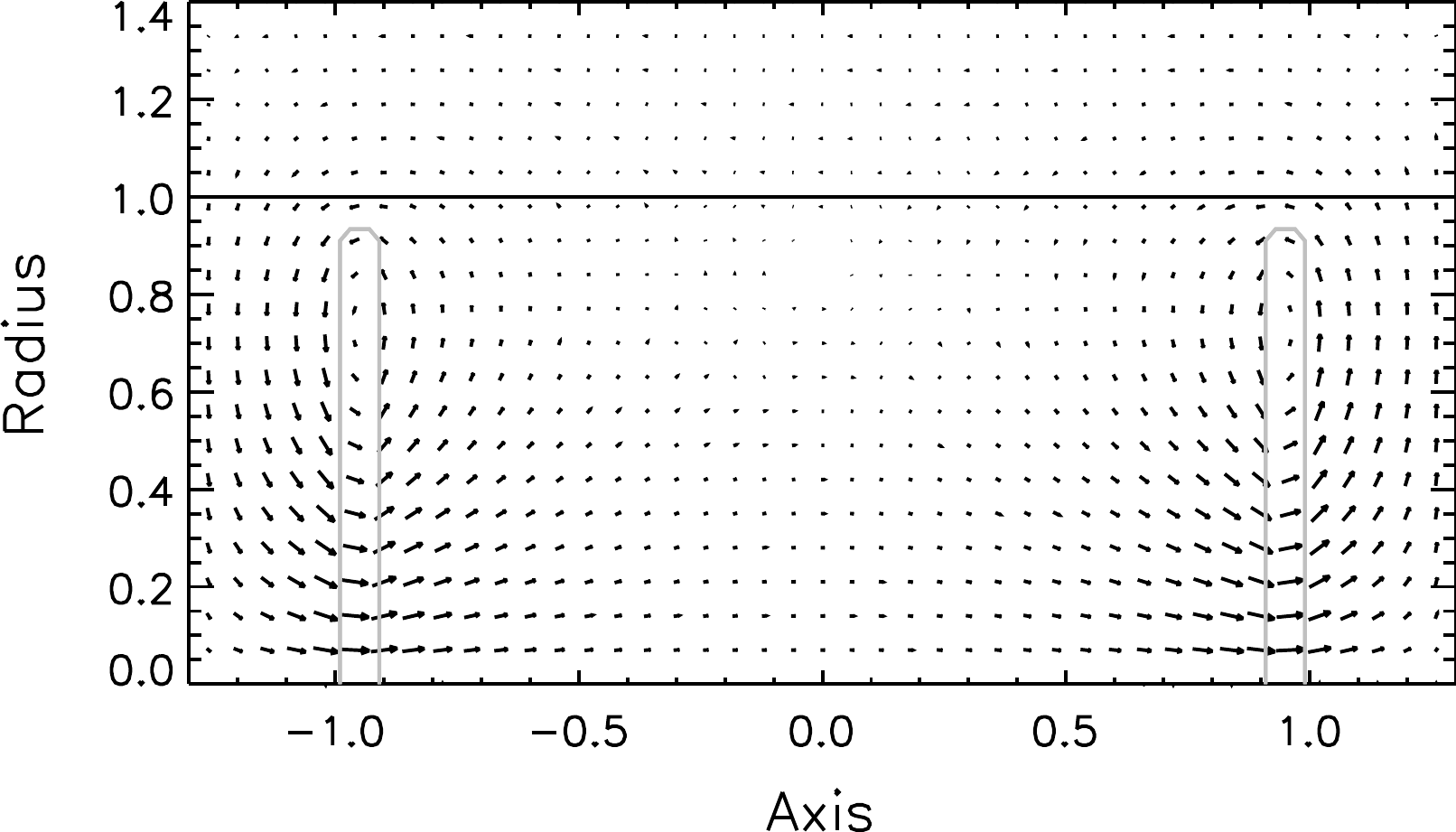}
\nolinebreak[4!]
\includegraphics[width=7cm]{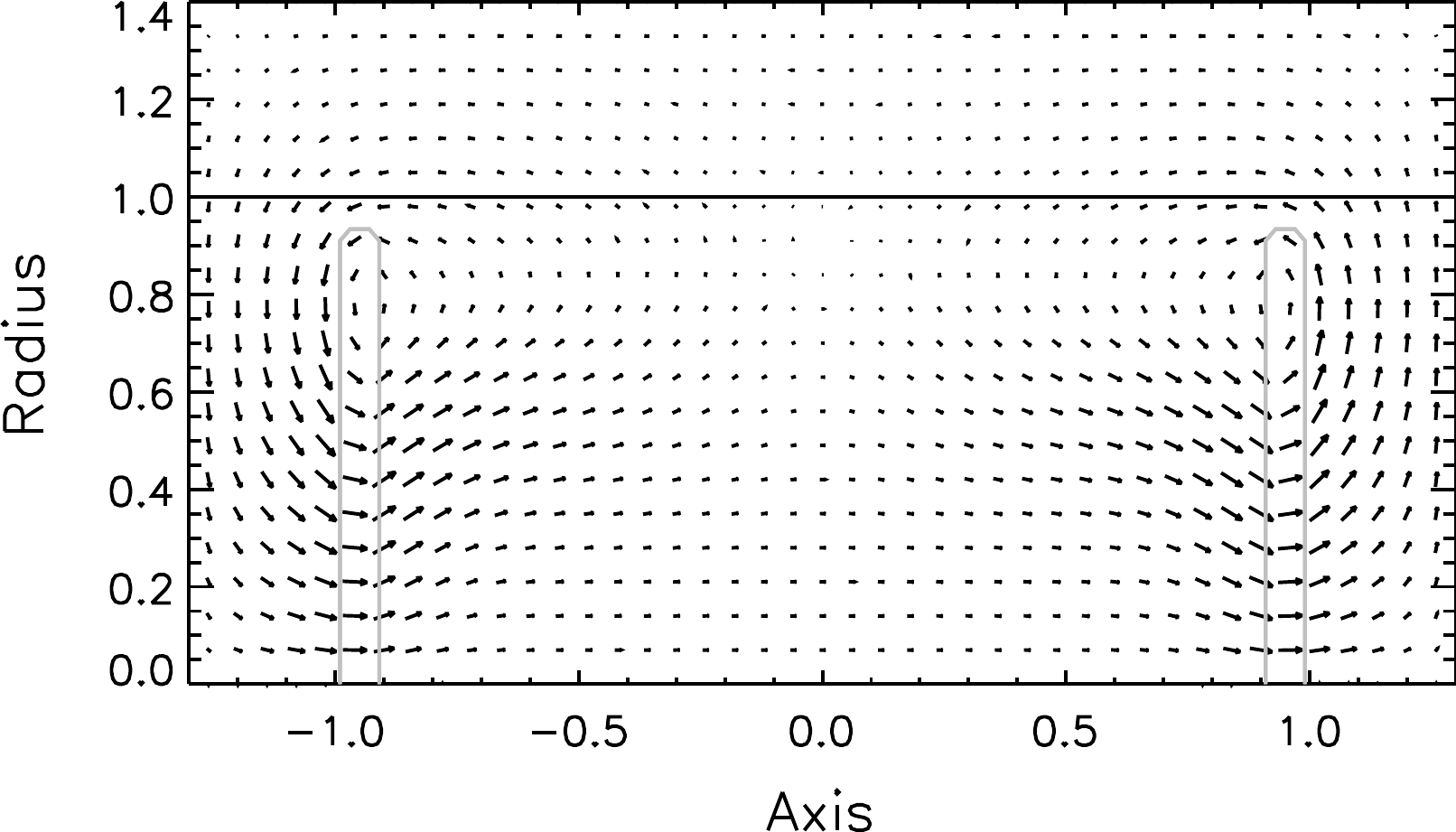}
\end{center}
\caption{
  Paramagnetic pumping at $\mur=60$.  Upper row: electromotive force
  (EMF)
  $\mathcal{E}^{\mu}=({\mu_0\mur\sigma})^{-1}\mur^{-1}{\nabla\mur}\times
  B_\vp\vec{e}_\vp$ at ${\rm{Rm}}=0$ (left) and ${\rm{Rm}}=30$
  (right).  Lower row: (poloidal) current density
  $\vec{j}=({\mu_0})^{-1}\nabla\times{\vec{B}}/{\mur}$ at
  ${\rm{Rm}}=0$ (left) and ${\rm{Rm}}=30$ (right).  The light grey lines
  show soft iron disks and the solid horizontal line shows the separation
  between the moving fluid and stagnant side layer.
  The azimuthal current is negligible. Note the close similarity between free
  decay (left column) and the case ${\rm{Rm}}=30$ (right column)
  illustrating the marginal impact of the fluid flow.
\label{fig::emf_current}} 
\end{figure}

The interaction between $\vec{V}^{\mu}\propto
\frac{\partial}{\partial z}\mur \vec{e}_z$ and the radial field
$B_r\vec{e}_r$ creates an azimuthal current $j_\vp\vec{e}_\vp$ at
the interface between the impeller disks and the fluid (see left panel
in~\fref{fig::emf_current_sketch}).
Since the impeller disks are thin, it is reasonable to assume that
the orientation and the amplitude of $B_r$ do not change across
the disks. This implies that the signs of the pumping term~(\ref{eq::pump}) at the back and
at the front side of the disks are opposite, which in turn implies that the
induced azimuthal currents mostly cancel each other and the overall azimuthal
current is close to zero.
This cancellation mechanism would not occur with thick disks. When the
impeller disks are thick, the mixed mode and the purely toroidal
mode have similar growth-rates as can be observed in the left panels
on figure~8 in \citename{2010GApFD.104..505G}
\citeyear{2010GApFD.104..505G} where the above phenomenon is
illustrated for two thicknesses of the impeller disks, $d=0.1$ and
$d=0.6$.

The behavior is very different concerning the EMF that
results from the interaction of the azimuthal component
$B_\vp\vec{e}_\vp$ with $\vec{V}^{\mu}$ (see right panel
in~\fref{fig::emf_current_sketch}).  
In this case the currents generated at the front and at the back of
the impeller disks add up and the EMF drives a poloidal current along
the surface of the disk which in turn re-enforces $B_\vp$.
Typical patterns of the EMF and current density from our numerical
simulations are shown in~\fref{fig::emf_current}. These graphics
confirm the presence of the poloidal current and confirm also that
the influence of the fluid flow is marginal.
\subsection{Simplified model for the toroidal $m0$-mode}
Our numerical results clearly indicate that the influence of the flow
on the toroidal axisymmetric mode is negligible and that this mode
is mostly localized inside the impeller disks. In order to better
understand the dynamics of the toroidal $m0$-mode, we 
consider an idealized disk-fluid model in free decay situation ($\Rm
= 0$).

Let us assume a disk of radius $1$, permeability
$\mu_r\gg 1$ and thickness $d$, sandwiched between two semi-infinite
cylindrical fluid regions with $\mur=1$. 
We further assume the boundary condition
$\vec{H}{\times}\vec{n}=0$ at the wall $r=1$. 
This simplifying assumption will allow us to find analytical solutions.  
We solve
\begin{eqnarray}
\mu_r  \gamma B_\vp  =  \Delta_* \, B_\vp, \quad \quad &&\quad  \quad  r< 1, \quad\quad | z| < d/2 \nonumber \\
\ \ \ \gamma B_\vp = \Delta_* \,B_\vp, \quad \quad &&\quad \quad r <
1, \quad\quad |z| > d/2 
\end{eqnarray}
where $\Delta_*=\frac{\partial^2}{\partial r^2}
+\frac{1}{r}\frac{\partial}{\partial r} 
+\frac{\partial^2}{\partial z^2} - \frac{1}{r^2}$. 
Note that the non-dimensionalization is done so that the reference
scale of the growth-rate is $(\sigma \mu_0 R^2)^{-1}$.  
The boundary condition is $B_\vp = 0$ at $r=1$, and the transmission
condition across the material interface is that $H_\vp$ and $E_r
= \partial_z H_\vp / \sigma$ be continuous at $z= \pm d/2$.

The numerical simulations show that the dominating eigenmodes are
symmetric with respect to the equatorial plane of the disk $z=0$. This
leads to the following ansatz   
\begin{eqnarray}
  B_\vp & = & A_1 \,  J_1 (kr) \,  \cos l_1 z, \ \quad \quad r< 1, \quad\quad  0< z < d/2  \nonumber \\
  B_\vp & = & A_2 \, J_1 (kr) \, e^{- L_2 z}, \quad \quad \quad r< 1,
  \quad\quad z > d/2, \nonumber 
\end{eqnarray}
where $J_1$ is the Bessel function of the first kind and
\begin{equation} 
  l_1 = \sqrt{-\gamma
    \mu_r -k^2}, \quad  \quad L_2 = \sqrt{k^2 +
    \gamma}. \label{kfun} 
\end{equation} 
The amplitudes $A_1$ and $A_2$ are arbitrary for the moment. 
The fields are obtained by symmetry for $z<0$. To ensure that the
solution decays at infinity (i.e. it remains bounded when $z\to
\pm \infty$), it is necessary that $l_1$ and $L_2$ be real. This
imposes the constraints
\begin{equation} 
  \mu_r > 1 \quad , \quad\gamma \in [-k^2, -k^2/\mu_r].
\end{equation} 
The boundary condition at $r=1$ implies that
$J_1(k)=0$, i.e. $k$ is a zero of $J_1$. We choose the first zero,
say $k_1$, so that the solution is composed of one radial cell only,
\begin{equation}
  k_1\approx 3.8317.
\end{equation}
This value specifically depends on the idealized boundary condition
that we have assumed at $r=1$; the effect of small deviations $k=(1\pm
0.1) k_1$ is considered further below.
Due to the assumed symmetry, we need to
impose the transmission conditions at the $z=d/2$ interface only: 
\begin{eqnarray}
\ \left ( \frac{1}{\mu_r} \ \cos \frac{l_1 d}{2} \right ) A_1 -
\left ( e^{-L_2 d/2} \right ) A_2 & = & 0, \nonumber
\\
\left(\frac{l_1}{\sigma \mu_r} \, \sin \frac{l_1 d}{2}\right)A_1 -
\left (\frac{L_2}{\sigma} \, e^{-L_2 d/2} \right ) A_2 & = & 0.
\end{eqnarray}
The determinant of the above linear system must be zero for a solution to exist,
\begin{equation} 
  L_2 \cos \frac{l_1 d}{2} - l_1 \sin \frac{l_1 d}{2} = 0.\label{eq::deter}
\end{equation}
Upon inserting the definitions of $l_1$ and $L_2$ from (\ref{kfun})
into this dispersion relation, we obtain an implicit nonlinear
equation for the growth-rate $\gamma$ as a function of $d$ and $\mur$.

\begin{figure} [b!]
\begin{center}
\includegraphics[width=8cm]{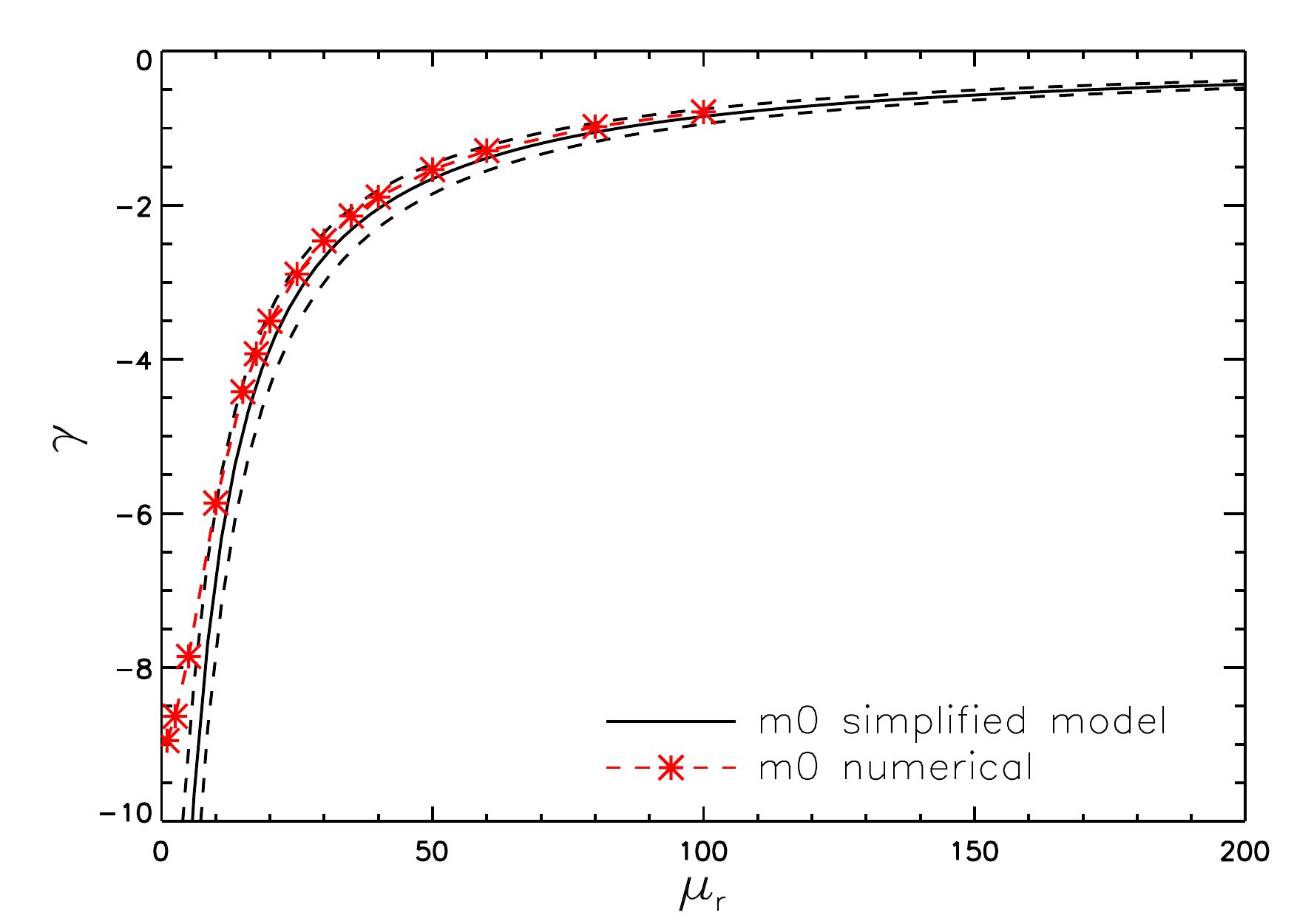}
\nolinebreak[4!]
\includegraphics[width=8cm]{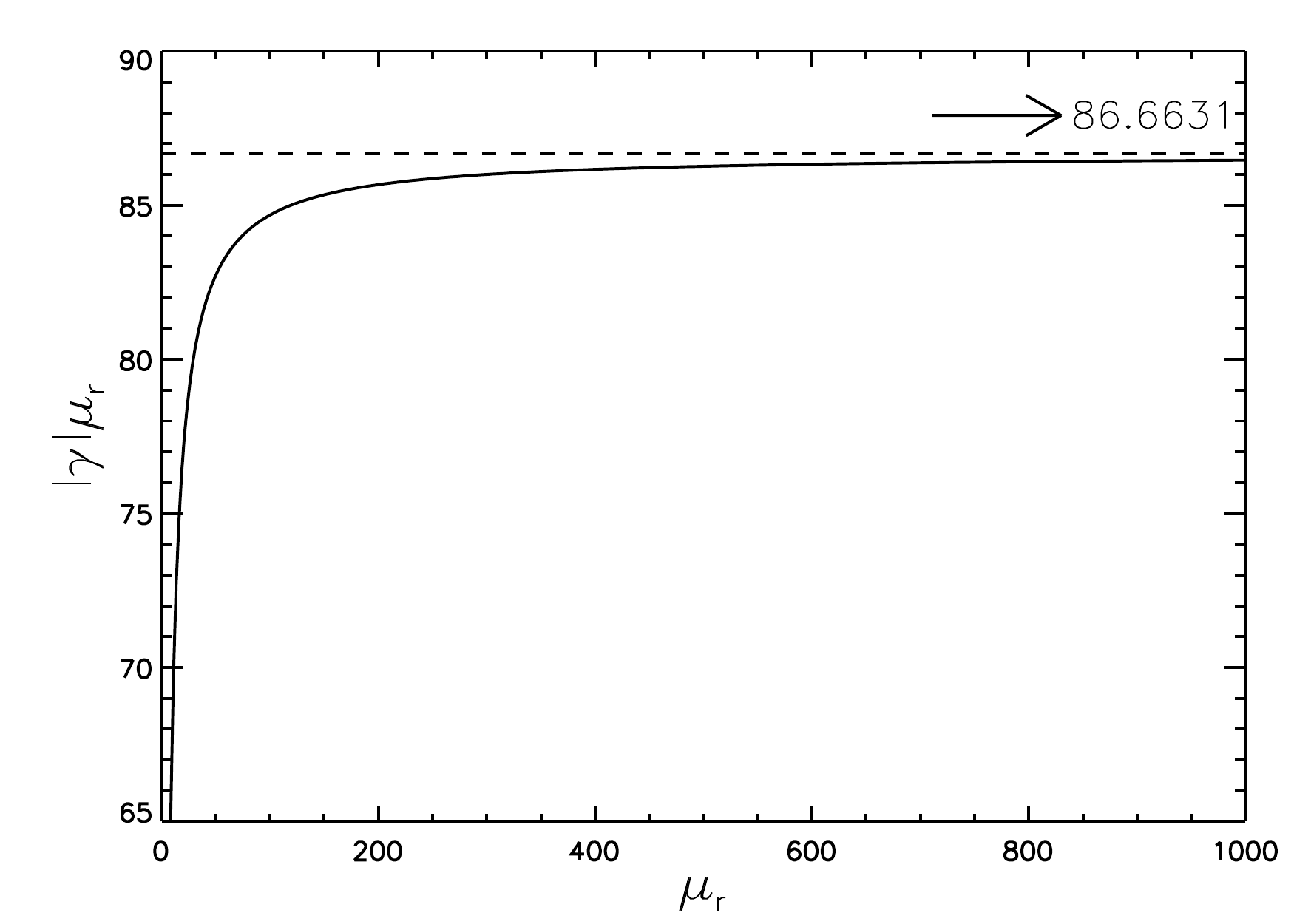}
\end{center}
\caption{
(left) Growth-rate $\gamma$ as a function of $\mu_r$ for the
  dominant axisymmetric toroidal eigenmode from the simplified
  model. $d=0.1$, $k=k_1$ and $k=(1\pm 0.1)
  k_1$. 
  (right) Using the proper time scale $\sigma \mu_r \mu_0
  R^2$ involving the permeability of the impeller disk, the growth-rate
  $\tilde{\gamma} = \gamma \mu_r$ reaches an asymptote at high
  $\mu_r$. 
  \label{fig4th}} 
\end{figure}

We show in the left panel of~\fref{fig4th} the graph of the function
$\gamma(\mur)$ deduced from~(\ref{eq::deter}) with
$d=0.1$ and $k=k_1=3.8317$. 
Two further analytical graphs computed with $k=(1\pm0.1)k_1$ 
show the very weak sensitivity of the growth rate on relaxing the
strict boundary condition $\vec{H}\times\vec{n}=0$ (e.g. by using an
outer shell of different conductivity).
We also show in this figure the numerical outcome for the 
growth-rate of the purely toroidal mode 
at $\rm{Rm}=0$ (see also
black solid line in~\fref{fig::gr_vs_mur_m0}). The agreement is quite
satisfactory and thereby confirms our analysis.
When representing $\gamma(\mur)$ in $\log$-$\log$ scale (not shown)
we see that $\gamma(\mur)$ scales like $\mu_r^{-1}$ in the limit
$\mu_r \rightarrow \infty$.  
Actually this behavior depends on the choice
that we have made for the non-dimensionalization.  
Involving the disks permeability for defining the new timescale
$(\sigma \mu_r \mu_0 R^2)$ instead of $(\sigma \mu_0 R^2)$,  
we obtain the rescaled growth-rate $\tilde{\gamma} = \gamma \mu_r$
shown in the right-panel of~\fref{fig4th}.
This representation
shows that the growth-rate $\tilde \gamma$ reaches a constant value
for very high permeabilities ($\tilde{\gamma}_\infty =
- 86.6631$). This observation immediately implies that the
following power law $\gamma \sim \tilde{\gamma}_\infty / \mur$ holds in
the original units when $\mur\to +\infty$.

In conclusion, the above simplified model explains why the growth-rate
of the purely toroidal mode goes to zero when $\mu_r \rightarrow
\infty$.  The model shows also that although the dominant purely
toroidal mode is localized to a very small volume, its decay time
determines the overall decay of the axisymmetric azimuthal magnetic
field.  Note finally that this mode would not be observed in numerical
simulations of VKS-dynamos that use the idealized boundary condition
$\vec{H}{\times}\vec{n}=0$ on the disk's surface (see
e.g.~\citeasnoun{2008giss}).
\section{Conclusions}
The aim of this paper is to study the influence of thin high
permeability disks on a VKS-like dynamo model. 
This work goes well beyond the study of \citeasnoun{2010GApFD.104..505G} in the
sense that we investigate thoroughly the axisymmetric mode and
present novel details on the scaling behavior of the dynamo
$m1$-mode.
The high permeability disks facilitate the occurrence of
non-axisymmetric dynamo action by enhancing the growth-rate of the
equatorial dynamo $m1$-mode. 
Compared to the idealized limit ($\mu_r\rightarrow \infty$) the
presence of a finite but high permeability material adds a small
supplementary damping effect and therefore slightly increases the dynamo
threshold.
We propose that the observed $\mu_r^{-1/2}$-scaling for the dynamo
threshold can be explained by a skin-effect
so that that the disk's role on the $m1$-mode is quite passive. 
Although the reduction of the magnetic Reynolds number is substantial
(from ${\rm{Rm}}^{\rm{c}}\approx 76$ at $\mur=1$ to
${\rm{Rm}}^{\rm{c}}\approx 54$ in the limit $\mur\rightarrow\infty$)
the spatial structure of the
$m1$-mode is hardly changed.

The effects of the high permeability of the impeller disks 
on the axisymmetric mode turn out to be more fundamental. 
In the presence of a mean flow the axisymmetric $m0$-modes
are split up in two separate families, one consisting of a purely
toroidal mode and one consisting of a mixed mode 
composed of a poloidal and a toroidal component. 
The growth-rate of the mixed $m0$-mode slightly increases with
$\rm{Rm}$ but is nearly independent of the disk
permeability.  
The growth-rate of the purely toroidal
$m0$-mode is not significantly influenced by the flow amplitude, but
it is considerably enhanced $\propto \mur^{-1}$ for large values of
$\mu_r$.  
This selective enhancement of the purely toroidal $m0$-mode can be
explained qualitatively by paramagnetic pumping. 
A simplified analytical model that interprets the purely toroidal mode
as a localized free decay solution confirms the scaling obtained in
the numerical simulations.  
This slowly decaying purely toroidal mode promoted through the
high permeability disks may play an important role in
axisymmetric dynamo action. However, in our simple axisymmetric set-up
no possibility for a closure of the dynamo cycle is provided  
since the poloidal component remains decoupled from the dominant toroidal
field so that such dynamo remains impossible
\cite{1933MNRAS..94...39C,1982GApFD..19..301H}.

Our study shows that 
the ideal boundary conditions $\vec{H}\times\vec{n}=0$ is indeed a
reasonable assumption for the $m1$-mode, but it is not appropriate for
the analysis of the toroidal $m0$-mode.
The purely toroidal $m0$-mode can be obtained only by explicitly
considering the internal permeability distribution and the corresponding
jump conditions at the fluid/disk interface.
This mode cannot be obtained numerically by simulations of VKS-like
dynamos that use either the idealized boundary condition
$\vec{H}{\times} \vec{n} = 0$ at the fluid/disk interface or the
thin-wall approximation from \citeasnoun{2010GApFD.104..207R}.

In conclusion, we have seen that the high (but finite) permeability in
the impeller disks is very important to promote axisymmetric
modes in our model and we suppose that it may
also play a  nontrivial role in the real VKS experiment.   
For example, in the presence of more complex
(non-axisymmetric) flows containing small scale turbulence modeled
by an $\alpha$-effect \cite{2008PhRvL.101j4501L,andregafd}, or in the
presence of non-axisymmetric permeability distributions
that resemble the soft-iron blades attached to the disks
\cite{PhysRevLett.104.044503}, 
the purely toroidal $m0$-modes can be coupled with poloidal field
components thus providing the required mechanism to close
the dynamo loop.

\ack{We acknowledge intense and fruitful discussions with J.-F. Pinton
and G. Verhille. 
AG is grateful to the assistence from T. Wondrak in
the implementation of the cublas library. 
Financial support from
Deutsche Forschungsgemeinschaft (DFG) in frame of the 
Collaborative Research Center (SFB) 609 is gratefully acknowledged. The
computations using SFEMaNS were carried out on IBM SP6 of IDRIS (project
0254).} 
\section*{References}
\bibliographystyle{jphysicsB}

\end{document}